\documentclass[aps,pre,twocolumn,showpacs,floatfix,groupedaddress,nofootinbib]{revtex4}

\usepackage{algorithm}
\usepackage{algpseudocode}
\usepackage{amsmath}
\usepackage{amsthm}
\usepackage{amssymb}
\usepackage{amscd}
\usepackage{bm}
\usepackage{latexsym}
\usepackage{mathtools}
\usepackage{comment}

\usepackage{graphicx}
\usepackage{epsfig}
\usepackage{epstopdf}
\usepackage{svg}
\usepackage[caption=false]{subfig}
\usepackage{placeins}
\usepackage{tabularx}

\usepackage{url}
\usepackage{hyperref}

\renewcommand{\vec}[1]{\mathbf{#1}}
\providecommand{\mat}[1]{\mathbf{#1}}

\providecommand{\config}[1]{\mathbf{#1}}

\providecommand{\norm}[1]{\Vert#1\Vert}

\providecommand{\config}{{\bf C}}
\providecommand{\eqref}[1]{(\ref{#1})}

\providecommand{\etal}{et al.\ }
\providecommand{\ie}{i.e., }
\providecommand{\eg}{e.g., }
\providecommand{\vanish}[1]{}

\newtheorem{theorem}{Theorem}
\newtheorem{conjecture}[theorem]{Conjecture}

\begin{document}
	
\title{Configuration spaces of hard spheres}

\author{O. B. Eri\c{c}ok}
\affiliation{Materials Science and Engineering, University of California, Davis, CA, 95616, USA.}

\author{K. Ganesan}
\affiliation{Materials Science and Engineering, University of California, Davis, CA, 95616, USA.}

\author{J. K. Mason}
\email{jkmason@ucdavis.edu}
\affiliation{Materials Science and Engineering, University of California, Davis, CA, 95616, USA.}

\begin{abstract}
Hard sphere systems are often used to model simple fluids. The configuration spaces of hard spheres in a three-dimensional torus modulo various symmetry groups are comparatively simple, and could provide valuable information about the nature of phase transitions. Specifically, the topological changes in the configuration space as a function of packing fraction have been conjectured to be related to the onset of first-order phase transitions. The critical configurations for one to twelve spheres are sampled using a Morse-theoretic approach, and are available in an online, interactive database. Explicit triangulations are constructed for the configuration spaces of the two sphere system, and their topological and geometric properties are studied. The critical configurations are found to be associated with geometric changes to the configuration space that connect previously distant regions and reduce the configuration space diameter as measured by the commute time and diffusion distances. The number of such critical configurations around the packing fraction of the solid-liquid phase transition increases exponentially with the number of spheres, suggesting that the onset of the first-order phase transition in the thermodynamic limit is associated with a discontinuity in the configuration space diameter.
\end{abstract}

\pacs{}

\maketitle

\section{Introduction}
\label{sec:introduction}
	
While statistical thermodynamics is generally considered to provide a satisfactory explanation of equilibrium states, the features of the configuration space that regulate the appearance of a phase transition are not yet entirely clear. Consider $n$ particles in a spatial domain $X$. Let $\Gamma_n$ be the phase space of the system such that a single point in $\Gamma_n$ completely specifies the microscopic state, \ie the positions and momenta of all the particles. Given an initial microstate, the system's time evolution corresponds to a continuous trajectory in the phase space. A thermodynamic property measurement effectively averages the value of this property along the portion of the trajectory corresponding to the measurement time interval. The ergodic hypothesis \cite{boltzmann1912vorlesungen} suggests that this time average can be replaced by a weighted average over the entire phase space if the time required to explore the phase space is substantially less than that of the measurement; the weight is a canonical measure that indicates the probability of observing a microstate at a generic instant in time. Finally, the Ehrenfest classification of phase transitions defines the transition order as the degree of the lowest derivative of the free energy that is discontinuous \cite{jaeger1998ehrenfest}. If ensemble averages are used to calculate these derivatives, and the canonical measure and the thermodynamic properties are smooth functions on the phase space, then from where do the discontinuities at a phase transition arise? If one objects that discontinuities should only be expected in the thermodynamic limit, there should nevertheless be incipient discontinuities in finite systems that develop into proper discontinuities in the limit. From where do these incipient discontinuities arise?

\begin{figure}
	\centering
	\includegraphics[width=0.7\columnwidth]{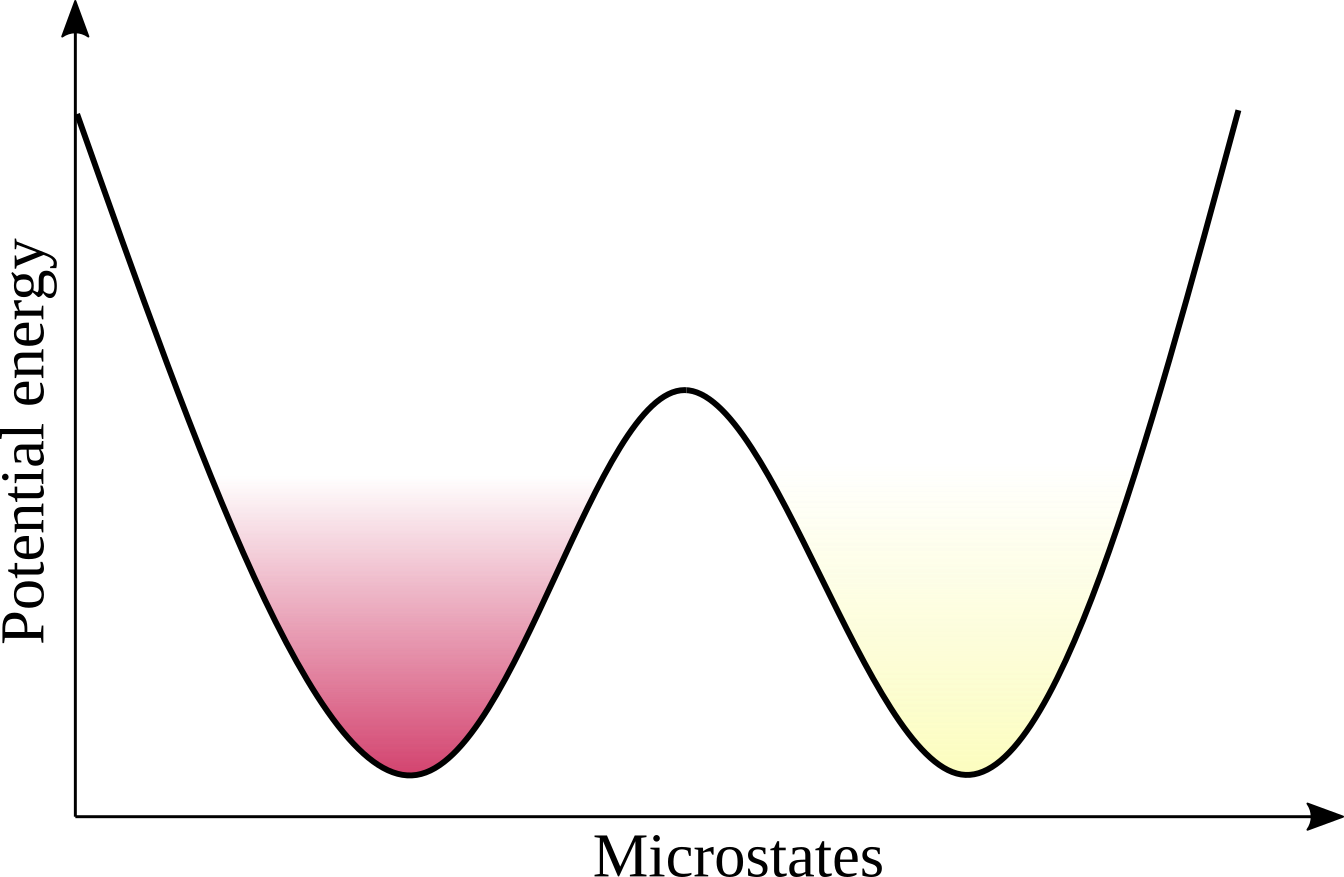}
	\caption{For internal energies below that of the barrier, the system is practically confined to a single local basin of attraction depending on the initial microstate. That is, ergodicity is broken.}
	\label{fig:microstates}
\end{figure}

The Topological Hypothesis \cite{caiani1997geometry,franzosi2004theorem} roughly suggests that the essence of the difficulty is the assumption that ergodicity holds close to a phase transition. That is, averages over the phase space should only be performed over the region that is effectively accessible from the initial microstate. If the extent of this accessible region changes discontinuously as a function of some thermodynamic variable, then the region over which the averages are computed would change discontinuously as well, and could result in discontinuous thermodynamic properties. Consider the example system shown in Fig.\ \ref{fig:microstates}. If the internal energy of the system is below the energy of the barrier, then the system is effectively confined to either the red or yellow basin of attraction depending on the initial conditions. Thermodynamic averages should then be computed over the region to which the system is confined, and increasing the internal energy to a level slightly above that of the barrier would roughly double the domain of integration. By contrast, the ergodic assumption suggests that the averages should be computed over both basins for all values of the internal energy, and is unreasonable from the standpoint of possible trajectories of the microstate.

The Topological Hypothesis more specifically asserts that discontinuous changes to the accessible part of the configuration space with increasing potential energy are associated with topological changes in the equipotential energy submanifolds; the configurations associated with these topological changes are known as critical points (a name that unfortunately conflicts with that for the end point of a phase equilibrium curve). Intuitively, the critical points could be the highest energy points along paths between two previously disconnected regions, or shortcuts between two previously distant regions. Franzosi \etal claimed to prove that a topological change in the equipotential energy submanifolds is a necessary condition for a phase transition to occur in systems with smooth, stable, confining, and short-range interactions \cite{franzosi2007topology1,franzosi2007topology2}. Kastner and Mehta \cite{kastner2011phase} subsequently studied a two-dimensional $\Phi^4$ system and observed a second-order phase transition occurring in an energy interval that didn't contain any critical points, disproving the claim. Gori \etal \cite{gori2018topological} responded by refining the hypothesis, suggesting that the phase transition in the $\Phi^4$ system is actually related to a divergence in the time required for the system to transition between two parts of the configuration space.

This literature does suggest that there is a relationship between the distribution of critical points and the onset of a phase transition, but a less direct relationship than initially supposed. Perhaps the natural signature of a phase transition is a discontinuity in the mixing time required for the system to adequately explore the configuration space, and such a discontinuity is generally (but not always) associated with topological changes and critical points. Slightly more formally:
\begin{conjecture}
\label{conj:phase_transition}
	A first-order phase transition occurs when there is a non-analyticity in the diameter (\ie the distance between the two most distant points) of the configuration space.
\end{conjecture}
\noindent This conjecture is deliberately vague with regard to the definition of the distance on the configuration space; if a phase transition is actually associated with a violent change to the configuration space geometry, then the non-analyticity should be relatively robust to the choice of distance. Reasonable choices should be related to geometric connectedness though, given the proposed relationship between a phase transition and the mixing time. The literature provides at least two distances on graphs that are suitable candidates: the commute time distance \cite{fouss2006novel,von2014hitting} and the diffusion distance \cite{coifman2006diffusion,talmon2013diffusion}. The former roughly measures the average number of steps taken by a random walker to go to a specified point and return to the starting point. The latter roughly measures the difference in concentration profiles of a species diffusing on the manifold when starting from Dirac delta distributions at two different points. Both are sensitive to the length and number of paths connecting any given pair of points, with small distances meaning there are a large number of short paths connecting them. 

The hard sphere system is one of the simplest systems that exhibits a phase transition, and it is often used as a prototype to model simple fluids \cite{dyre2016simple}. It is governed by the hard sphere potential, \ie the energy is infinite if the spheres overlap and zero otherwise. Let $\eta$ be the packing fraction, defined as the fraction of volume occupied by spheres. Alder and Wainwright were the first to identify different phases in the two-dimensional hard disk system as a function of $\eta$ \cite{alder1962phase}, and continued their work on hard sphere systems \cite{wood1957preliminary,alder1957phase,alder1968studies}. Isobe and Krauth \cite{isobe2015hard} studied the crystallization and melting of a system of a million hard spheres with event-chain Monte Carlo, local Monte Carlo and event-driven molecular dynamics simulations, and observed a liquid-solid phase coexistence when $0.490<\eta<0.548$. Recently, Pieprzyk \etal \cite{pieprzyk2019thermodynamic} studied the thermodynamic and dynamical properties of hard spheres using molecular dynamics and observed phase coexistence in the same interval.

Previously, Carlsson \etal \cite{carlsson2012computational} studied the configuration space of hard disk systems in a square, and found the critical points for five disks by regularizing the potential energy surface. Baryshnikov \etal \cite{baryshnikov2014min} then started to develop a min-type Morse theory for the configuration spaces of hard spheres and proved that the critical points correspond to mechanically-balanced configurations. Ritchey \cite{ritcheyphd} studied the configuration spaces of hard disks in the hexagonal torus for $n=1 \dots 12$ disks, created a database of critical points, and identified the wallpaper groups of each critical point. Recently, Eri\c{c}ok and Mason \cite{ericok2021quotient} proposed descriptors and distances on the configuration spaces of hard disks in the square and hexagonal toruses modulo specified symmetry groups, and used this to construct explicit triangulations of the quotient spaces.

The purpose of this work is to develop a procedure to practically test Conjecture \ref{conj:phase_transition} using the distances mentioned above for hard sphere systems. Section \ref{sec:preliminaries} introduces various concepts and definitions that are relevant to the configuration spaces of hard spheres. As far as the authors know, the critical points of the configuration spaces of hard spheres have not been tabulated before. Section \ref{sec:critical_configuration} describes the known critical configurations of systems with $n = 1 \dots 12$ hard spheres on a rhombic dodecahedron with periodic boundary conditions, and the relevance of the distribution of critical points to the phase coexistence interval. The distances and descriptors proposed in Ref.\ \cite{ericok2021quotient} have been extended to the hard sphere systems. Section \ref{sec:configuration_spaces} gives approximate triangulations of the configuration spaces for small numbers of spheres, and explores the relationship of the critical configurations to the space geometry. Finally, Section \ref{sec:diffusion_distance} reviews these results in the context of the above conjecture, particularly for the configuration spaces of two hard spheres.

\section{Preliminaries}
\label{sec:preliminaries}

\subsection{Tautological function}
\label{subsec:tautological function}

The rhombic dodecahedron (RD) with periodic boundary conditions (discussed further in App.\ \ref{sec:app_3-torus}) is the domain of all the hard sphere configurations considered here. A hard sphere configuration is specified by the locations of the sphere centers and a sphere radius. Since the location of a single sphere center is equivalent to a point in a $3$-dimensional torus $\mathbb{T}^3$ (i.e., the RD with periodic boundary conditions), the configuration space of $n$ sphere centers is the product space of $n$ toruses, or $\Lambda(n)$. A real-valued function $\tau: \Lambda(n) \rightarrow \mathbb{R}$ called the tautological function is defined as the largest radius the spheres could all have without overlapping given the locations of the sphere centers, or
\begin{equation*}
\tau = \min\limits_{\substack{1 \leq i < j \leq n}} {r_{ij}}
\label{eq:tautological_function}
\end{equation*}
where $r_{ij}$ is half the shortest distance between the centers of spheres $i$ and $j$. The configuration space of $n$ hard spheres of radius $\rho$ is then more precisely defined as
\begin{equation}
\Gamma(n,\rho) = \tau^{-1}([\rho,\infty)),
\label{eq:configuration_space}
\end{equation}
or the subset of $\Lambda(n)$ such that the smallest distance between any pair of sphere centers is at least $2 \rho$.

\subsection{Morse theory}
\label{subsec:morse_theory}

Equation \ref{eq:configuration_space} shows that the configuration spaces of hard spheres can be written as the $\rho$-superlevel sets of $\tau$. This suggests that a Morse-type theory can be used to study the topological changes of $\Gamma$ as a function of $\rho$. Classical Morse theory \cite{matsumoto2002introduction,milnor2016morse} relates the topology of a smooth manifold $M$ to the critical points of a generic differentiable function $f$ defined on $M$, where a critical point is a point where the gradient $\nabla f$ vanishes. The value of the function at a critical point is referred to as the critical value, and the number of negative eigenvalues of the Hessian matrix at a critical point is referred to as the critical index. Intuitively, the critical index indicates the number of independent collective motions that allow $f$ to decrease to second order. This paper exclusively uses the term ``critical point'' to indicate ones in a Morse-theoretic setting, and not the end points of phase equilibrium curves.

Morse theory indicates that the topology of the superlevel sets of a manifold can only change at a critical point, though not all critical points are associated with a topology change. A critical point is said to be degenerate if the Hessian matrix is singular at that point; according to classical Morse theory, a non-degenerate critical point is always associated with a topology change, but the behavior of the function at a degenerate critical points is unpredictable. For example, $x = 0$ is a critical point of both $y = x^2$ and $y = x^3$, is a non-degenerate critical point in the first and a degenerate one in the second, and there is an associated topology change in the superlevel sets of only the first.

Let $M^a = f^{-1}(\left[a, \infty\right))$ denote the $a$-superlevel set of $M$, or the set of all points whose function value is at least $a$. The first theorem of Morse theory states that $M^a$ and $M^b$ are topologically equivalent to each other if the interval $[a, b]$ doesn't contain any critical value. If it contains the critical value of an index-$p$ critical point, then the topology of $M^b$ is equivalent to the topology of $M^a$ with a $p$-handle attached (a $d$-dimensional $p$-handle is defined as the product of a $p$-dimensional disk and a $(d-p)$-dimensional disk). In three dimensions, $0$- and $3$-handles are solid balls whereas $1$- and $2$-handles are solid cylinders, though they differ by the boundary along which they are attached.

\begin{figure}
	\centering
	\includegraphics[width=0.85\columnwidth]{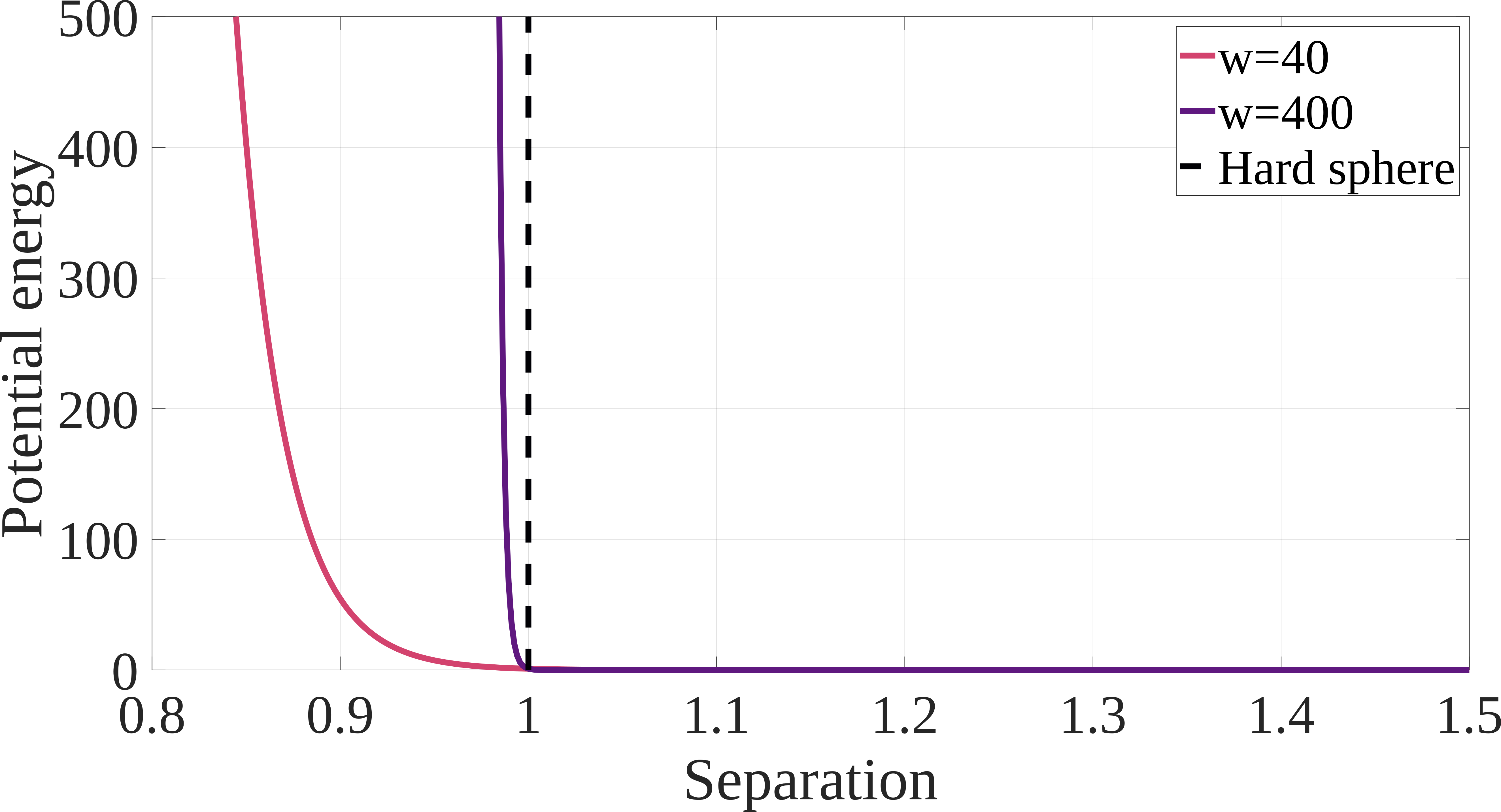}
	\caption{The hard sphere potential (black dashed line) is infinite if the separation of two sphere centers is less than the sum of their radii and zero otherwise. Observe that $E$ converges to the hard sphere potential as $w \to \infty$.}
	\label{fig:auxillary_energy}
\end{figure}

The difficulty with using the functions $\tau$ or the hard sphere potential in this context is that they are not differentiable. This is resolved by regularizing the hard sphere potential, resulting in an infinitely-differentiable potential energy. More specifically, the hard sphere potential is replaced by an exponential repulsion where $\rho$ and $w$ govern the onset and the strength of the repulsion, respectively:
\begin{equation*}
E = \sum_{i} \sum_{j > i} \exp[-w (r_{ij} - \rho)].
\end{equation*}
This can be seen to converge to the hard sphere potential in the limit of large $w$ in Fig.\ \ref{fig:auxillary_energy}. Moreover, the critical points of $E$ can be related to those of $\tau$ (despite $\tau$ not being differentiable) in the following way. Observe that the radius $\rho$ of the spheres could be increased by as much as $\tau - \rho = \min_{i < j}(r_{ij} - \rho)$ without the spheres overlapping. The soft-min approximation replaces the minimum by the logarithm of a sum of exponentials, or
\begin{equation}
\tau - \rho \approx -\frac{1}{w} \log \Big\{\sum_{i} \sum_{j > i} \exp[-w (r_{ij} - \rho)] \Big\},
\label{eq:deriv_soft_min}
\end{equation}
where the sum is generically dominated by a single term corresponding to the minimum in the limit of large $w$. Observe that the critical points of the right side of Eq.\ \ref{eq:deriv_soft_min} are the same as those of $E$ for any finite value of $w$ since the logarithm and multiplication by a constant are strictly monotone transformations, and the critical points of the left hand side are the same as those of $\tau$ since $\rho$ is a constant. Equation \ref{eq:deriv_soft_min} then suggests that the critical points of $\tau$ be defined to be those of $E$ in the large $w$ limit.

The critical index is calculated as the number of negative eigenvalues of the Hessian matrix of $E$ evaluated at the critical configuration for large but finite $w$. Ritchey \cite{ritcheyphd} provided an alternative definition of the index suitable for min-type functions that coincides with the above definition for every known non-degenerate critical point. This is discussed further in Sec.\ \ref{sec:critical_configuration} below.

\subsection{Filtrations of spaces}
\label{subsec:filtrations_of_spaces}

Given that our concern is with the geometry of the configuration space, it is convenient to construct a triangulation that represents this geometry explicitly. This is done by first constructing a triangulation of the configuration space of points $\Lambda(n)$ as a simplicial complex (a $k$-simplex is the analogue of a triangle in dimension $k$; further details are given in App.\ \ref{sec:app_simplicial_complex}). Each simplex of the triangulation is associated with a value of $\tau$ equal to the smallest value of any of the bounding vertices. A triangulation of the configuration space of hard disks $\Gamma(n, \rho)$ is then constructed as the subcomplex containing all simplices with associated values of $\tau$ greater than or equal to $\rho$, i.e., all triangles corresponding to locations of the sphere centers such that spheres of radius $\rho$ do not overlap. This allows the geometry of $\Gamma(n, \rho)$ to be monitored for changes as a function of $\rho$. For brevity, $\Gamma(n, \rho)$ is said to be a filtration of $\Lambda(n)$ by $\tau$. 

Changing the filtration value not only affects the triangulations of the spaces, but their topological invariants as well. One of these invariants is the so-called merge tree associated to a function $f:M \rightarrow R$ \cite{morozov2013distributed}. Intuitively, a merge tree tracks the way components merge or separate as the sublevel (or superlevel) value changes. Formally, two points $x$ and $y$ are considered equivalent if they belong to the same level set $f(x) = f(y) = \ell$ and belong to the same component of the sublevel set $f^{-1}((-\infty, \ell])$; quotienting $M$ by this equivalence relation gives the merge tree. There is a very similar object in the chemical physics literature known as a disconnectivity graph \cite{becker1997topology}. The merge tree can also be thought as a tree whose leaf nodes are minima (index-$0$), internal nodes are saddle points (index-$1$) and the root vertex is the global maximum \cite{morozov2013distributed}.

By hypothesis, the mixing time of a thermodynamic system depends on the connectivity of the space and the volumes of the accessible regions. This suggests that changes to the accessible domain and the expected time to explore the domain be analyzed by a distance function that is sensitive to variations in the length and number of paths connecting pairs of points. Two such distances on graphs are the commute time distance \cite{fouss2006novel,von2014hitting} and the diffusion distance \cite{coifman2006diffusion,talmon2013diffusion}. Given a weighted graph $G$, the average commute time distance roughly measures the average number of steps necessary for a random walker starting from vertex $i$ to reach vertex $j$ and return to $i$. The diffusion distance instead roughly measures the difference in concentration profiles of a species diffusing on the manifold when starting from Dirac delta functions at vertices $i$ and $j$. Both the commute time distance and the diffusion distance decrease with decreasing length of paths and increasing number of paths connecting vertices $i$ and $j$. These distances are briefly defined in App.\ \ref{sec:app_graph_distances}.

The diameter of a space is defined as the commute time or diffusion distance between the two most distant points, and is expected to increase with the mixing time of a thermodynamic system. The diffusion distance in particular is closely related to the mixing time, roughly defined as the time required for the probability distribution of microstates to converge to the Boltzmann distribution (setting aside Loschmidt's paradox) and for the preparation protocol to be forgotten. As such, any discontinuities in the mixing time as a function of $\rho$ should coincide with discontinuities in the diameter of $\Gamma(\rho, n)$ as measured by the commute time and diffusion distances. While both distances require the space to have the property that a path connects each pair of points, the distances can still be calculated separately for each component of a disconnected space.

As with the example in App.\ \ref{sec:app_graph_distances}, discontinuous jumps in the diameter are expected to occur when two or more previously disconnected components merge, and therefore to be associated with one or more critical configurations. This motivates first identifying critical points of the configuration spaces for $n = 1 \dots 12$ hard spheres to predict the radius intervals where geometric changes will occur. After analyzing the properties of these critical points, explicit triangulations for $n = 2$ spheres are constructed to verify that the appearance of critical points can substantially affect the geometric properties (\ie the diameter) in the expected way. If so, then it is reasonable to conjecture that the same could happen for systems with $n > 2$ spheres, or even in the thermodynamic limit.

\section{Critical Configurations}
\label{sec:critical_configuration} 

\begin{figure*}
	\centering
	\includegraphics[width=1.0\textwidth]{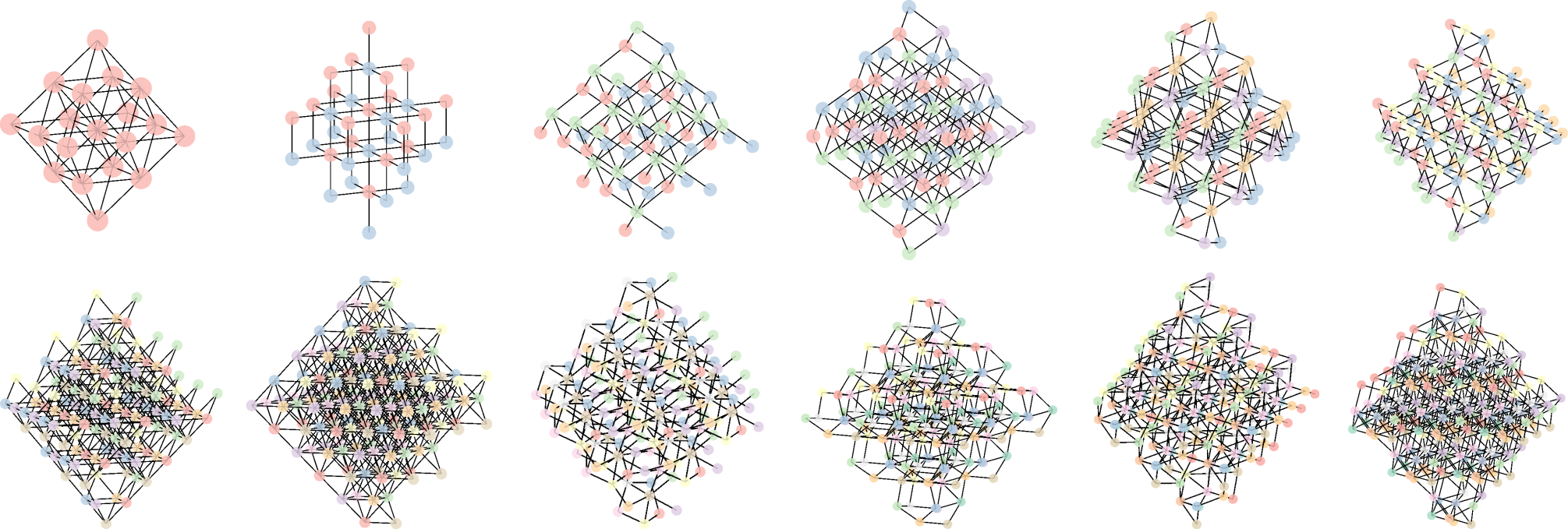}
	\caption{The densest configurations found for $n=1 \dots 12$. The rest of the critical configurations can be visualized using the interactive database described in the text. Some of the critical configurations correspond to the well-known crystal structures, e.g., the first and the eighth are face-centered cubic, the second is simple cubic and the fourth is body-centered cubic.}
	\label{fig:densest_configs}
\end{figure*}

As explained in Sec.\ \ref{subsec:morse_theory}, the critical points of $\tau$ are identified with those of $E$ in the limit of large $w$. The critical points of $E$ are defined by the condition $\nabla E = 0$, and are therefore minima of the nonnegative function $\lvert \nabla E \rvert ^2$. The locations of the critical points can therefore be sampled by repeatedly initializing a search with a random initial configuration and using any standard minimization algorithm (\eg conjugate gradient) with $\lvert \nabla E \rvert ^2$ as the objective function. While this does not provide an exhaustive enumeration of critical points, the results below show that tens of thousands of distinct critical points have already been identified in this way. It should be mentioned that there is an additional complication when $n < 9$, since for small values of $n$ it is possible for a sphere to be in contact with multiple periodic images of another sphere. The consequence of this is that all of the spheres in all $19$ images of the fundamental cell shown in Fig.\ \ref{fig:rhombic_dodecahedron} need to be considered in such situations. Where $n \ge 9$, multiple contacts are disallowed geometrically and only the spheres in the fundamental cell need to be considered.

Ritchey \cite{ritcheyphd} suggested that the critical points could be grouped into equivalence classes related by symmetry operations, including rigid translations, permutations of the sphere labels, and symmetries of the tessellation. An interactive database of all known critical points has been created for $n = 1 \dots 12$ spheres in the RD using a Colab Notebook and is freely available online\footnote{The database of critical points is available at \url{https://github.com/burakericok/hard_sphere_crits}.}. The densest known configurations for $n=1 \dots 12$ are shown in Fig.\ \ref{fig:densest_configs}.

A critical configuration is associated with a vector $\config{x}$ of coordinates of the sphere centers, an adjacency matrix $\mat{A}$, a sphere radius $\rho$, and a critical index $p$. For a given labeling of the spheres, $\mat{A}$ is an integer matrix where the element $a_{ij}$ is the number of contacts between spheres $i$ and $j$; the main diagonal is always zero since a sphere cannot be in contact with itself (except for $n = 1$). A sphere is called a rattler if it is not in contact with any other spheres. 

This description of a critical point is clearly not the same for all elements of a given equivalence class, \eg a translation adds a structured vector to $\config{x}$, and a permutation reorders the rows and columns of $\config{x}$ and $\mat{A}$. This is handled by using the program nauty \cite{mckay2014practical} to construct a canonical labeling of the spheres that is the same for every configuration in a given equivalence class, and translating the first sphere to the origin. After applying the canonical labeling, one can decide whether a candidate critical point belongs to an equivalence class that has already been observed on the basis of $\mat{A}$, $\rho$ and $p$.

Previously, Baryshnikov \etal \cite{baryshnikov2014min} showed that the critical points of the configuration spaces of hard spheres are precisely the mechanically-balanced configurations. Let $\vec{e}_{ij}$ be the vector pointing from the center of sphere $i$ to that of sphere $j$, and $w_{ij}$ be the force along or weight on this edge of the contact graph. A configuration is said to be mechanically balanced if there is a nontrivial set of repulsive forces along the edges such that the sum of forces on every sphere vanishes; that is, if there is a nontrivial set of non-negative weights such that $\sum_{j \sim i} w_{ij}\vec{e}_{ij} = 0$ for every fixed vertex $i$, where the sum is performed over all vertices adjacent to $i$. A critical configuration is said to be non-degenerate if, for each edge, there exists at least one mechanically-balanced set of weights such that the weight on that edge is positive.

\begin{figure}
	\centering
	\includegraphics[width=0.9\columnwidth]{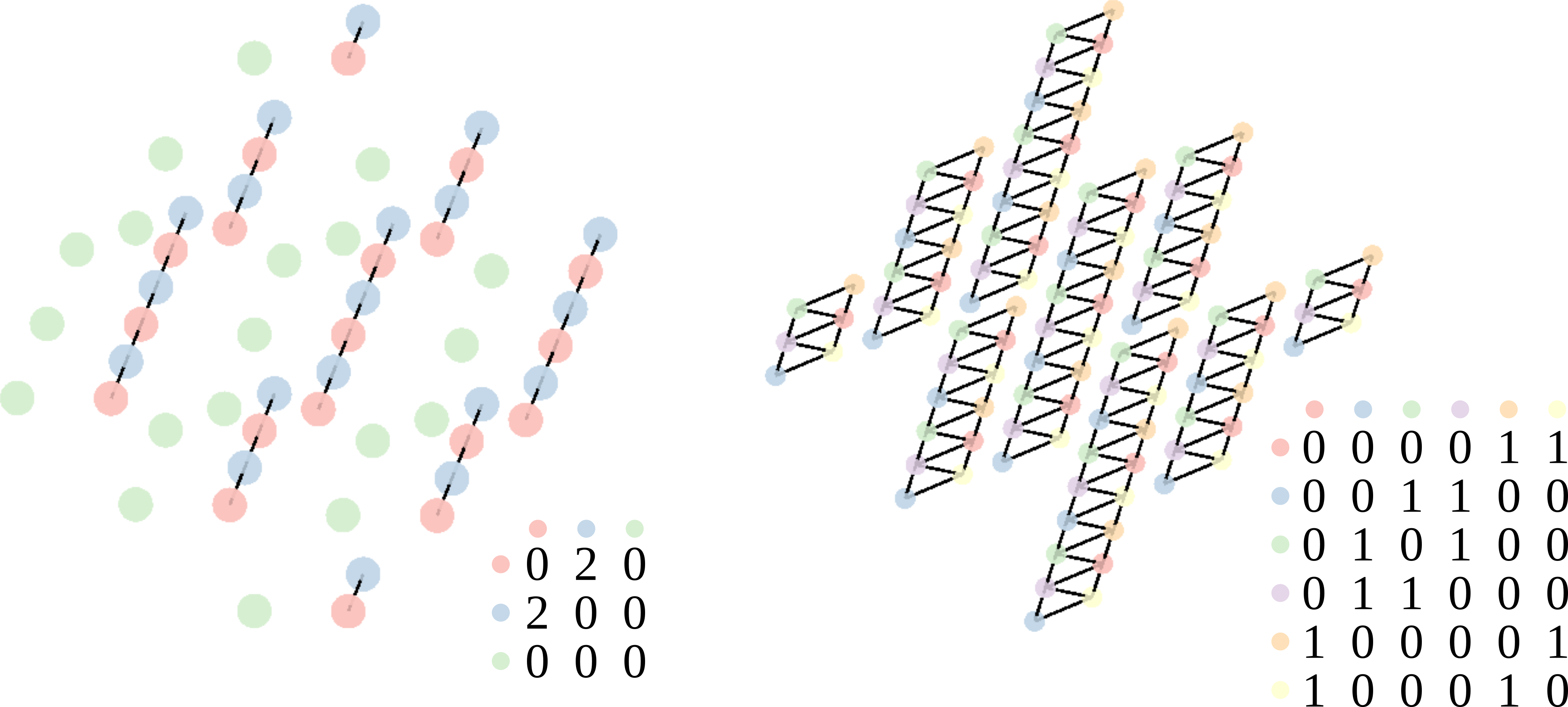}
	\caption{Examples of critical configurations for $n = 3$ (left) and $n = 6$ (right), with adjacency matrices indicated in the insets. The left one is a non-degenerate index-$2$ critical point with $\rho = 0.25$, whereas the right one is a degenerate index-$8$ critical point with $\rho = 0.167$. Observe that the edges between adjacent columns on the right are degenerate, and these edges are not reflected in the adjacency matrix.}
	\label{fig:crit_examples}
\end{figure}

\begin{figure}
	\centering
	\includegraphics[width=0.85\columnwidth]{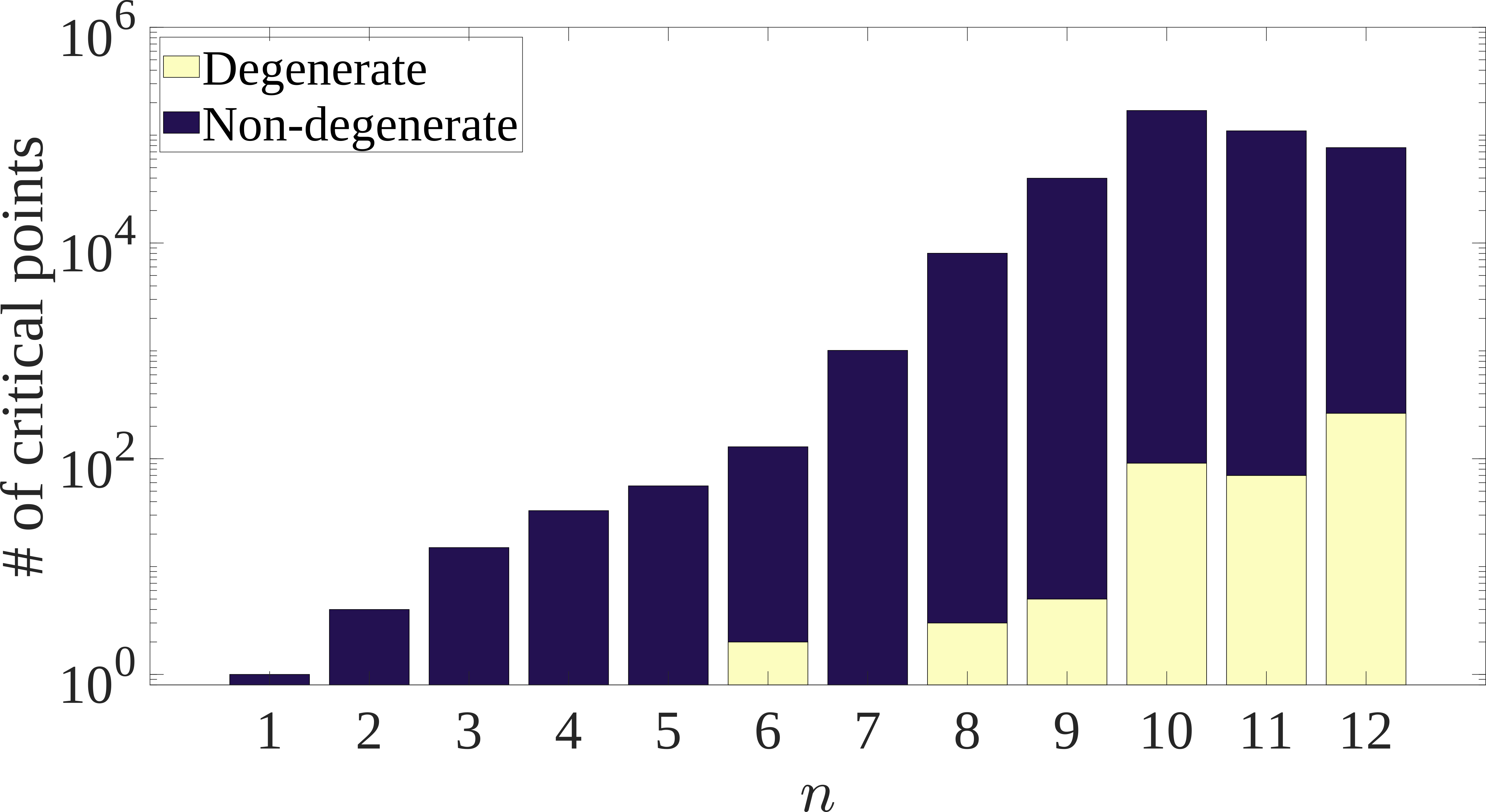}
	\caption{The number of equivalence classes of non-degenerate and degenerate critical points found by the algorithm described in Sec.\ \ref{sec:critical_configuration} for ten million random initial configurations for $n=1 \dots 10$ and three million for $n=11 \dots 12$.}
	\label{fig:nondegen_vs_degen}
\end{figure}

Figure \ref{fig:crit_examples} provides several example critical points to illustrate these definitions. The critical point on the left is a non-degenerate index-$2$ critical point with $n = 3$ spheres and $\rho = 0.25$, and the green sphere is a rattler since it has no contacts with the other spheres. The reason this is an index-$2$ critical point is that there are two independent directions that the blue sphere can be moved relative to the red one while increasing the radius to second order; any possible motions of the red sphere relative to the blue one can be made equivalent to these by a translation. The critical point on the right is an index-$8$ critical point with $n = 6$ spheres and $\rho = 0.1667$. The edges between two adjacent columns are all degenerate edges; it is not possible to assign positive weights to these edges without breaking mechanical balance. As Fig.\ \ref{fig:nondegen_vs_degen} shows, the majority of the critical points already found are non-degenerate. It is also significant that, even though the algorithm used to generate Fig.\ \ref{fig:nondegen_vs_degen} does not identify all critical points, the number of known critical points seems to increase exponentially in $n$ on the basis of the well-sampled interval $n = 6\dots10$.

\begin{figure}
	\centering
	\includegraphics[width=0.85\columnwidth]{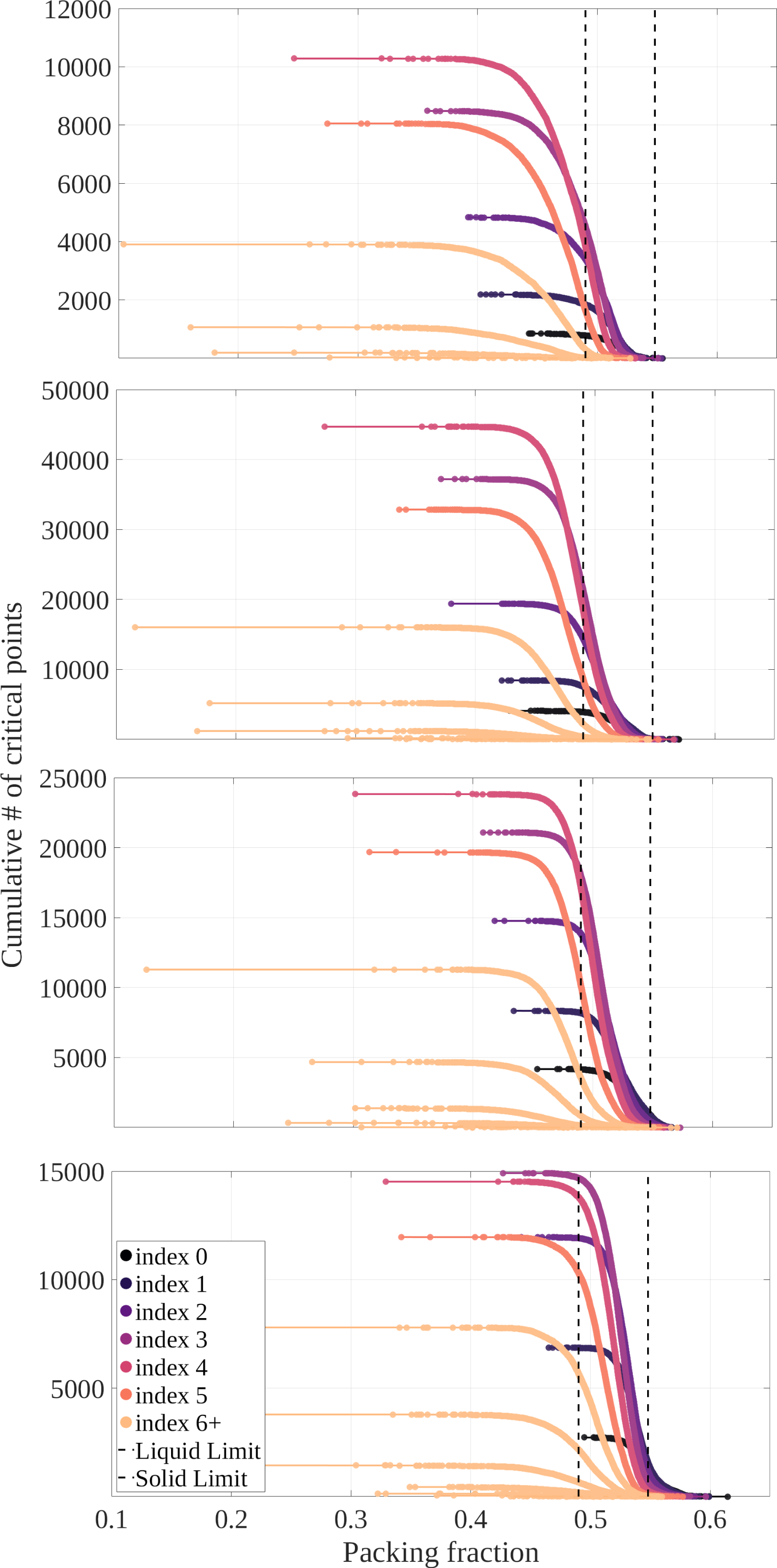}
	\caption{Distribution of equivalence classes of critical points for $n = 9 \dots 12$ (from top to bottom) as a function of index and packing fraction. The critical points are sorted by packing fraction, and the cumulative distributions are constructed for decreasing packing fraction (\ie right to left).}
	\label{fig:distributions}
\end{figure}

If the distribution of critical points is indeed related to the hard sphere phase transition, then it is reasonable to suppose that there will be a significant number of critical points at densities close to the liquid and solid limits of the infinite system, even when the systems are finite. Figure \ref{fig:distributions} shows the distributions of the critical points for $n = 9 \dots 12$ spheres as a function of their index and packing fraction $\eta$ (the fraction of volume occupied by spheres). Almost all of the critical points occur around the coexistence interval in each case, with the lower index critical points slightly below the solid limit and the higher index critical points closer to the liquid limit. Moreover, the distributions of the critical points change with increasing $n$, becoming more concentrated and shifting closer to the solid side of the coexistence interval. While this could be a result of biased sampling of the critical points for $11$ and $12$ spheres, closer study of the distributions for $9$ and $10$ spheres suggests that the distribution of lower index critical points either remains centered in the middle of the coexistence interval or narrows and shifts toward the solid limit with increasing $n$. Further observe that the volume of the configuration space and the number of critical points both increase exponentially with $n$ (Fig.\ \ref{fig:nondegen_vs_degen}), suggesting that the density of critical points in the configuration space does not change as rapidly as either one. If these two trends hold for all $n$, then any involvement of the critical point distribution in the appearance of a phase transition in the thermodynamic limit would likely remain unchanged even for systems with small $n$. This is the line of reasoning that motivates the detailed study of the $n = 2$ sphere system in the following sections.

\section{Configuration Spaces}
\label{sec:configuration_spaces}

The hard sphere system defined in Eq.\ \ref{eq:configuration_space} has a variety of symmetries, or physical operations that make multiple points in $\Gamma$ physically indistinguishable. These include continuous translations, permutations of the sphere labels, and the symmetries of the tessellation. For example, exchanging the labels of two spheres changes the point in $\Gamma$ but does not change the physical configuration. This phenomenon is typically handled by quotienting the configuration space by the appropriate symmetry group to identify each group of configurations related by symmetry operations to a single point. For example, the use of correct Boltzmann counting in statistical mechanics is generally regarded as being equivalent to quotienting the configuration space by permutations of the particle labels \cite{landau2013statistical}, and using the center of mass coordinates of a mechanical system effectively quotients the configuration space by translations \cite{denman1965invariance}.  The symmetry groups $\mathcal{S}_i$ and the quotient spaces $\Gamma / \mathcal{S}_i$ considered in this work are indicated in Table \ref{table:symmetries}. 
	
Ericok and Mason \cite{ericok2021quotient} proposed two methods to sample points in the quotients of the configuration spaces of hard disks by the desired symmetries. The first proposes an infinite set of real-valued descriptors that are invariant to the desired symmetries, mapping the full configuration space into an infinite-dimensional Euclidean space. The isometric feature embedding (ISOMAP) \cite{Tenenbaum2319} algorithm is used to reduce the dimension of the image as much as possible while preserving the local geometry as defined by the edge lengths in the $k$-nearest neighbor graphs.

\begin{table}
	\centering
	\begin{tabular}{c c} 
		Space     & Symmetries \\ [0.5ex] 
		\hline\hline
		$\Gamma/\mathcal{S}_1$ & $\mathcal{T}$\\
		$\Gamma/\mathcal{S}_2$ & $\mathcal{T} \bigcup \mathcal{P} \bigcup \mathcal{I} \bigcup \mathcal{L}$\\[1ex]
	\end{tabular}
	\caption{The spaces considered in this work. $\mathcal{T}$, $\mathcal{P}$, $\mathcal{I}$ and $\mathcal{L}$ are the groups of translations, permutations, inversions, and proper symmetries of the tiling.}
	\label{table:symmetries}
\end{table}

The second method uses a distance function $d_{\Lambda}(\config{p}, \config{q}) = \sum_{i = 1}^{n} \lVert \vec{p}_i - \vec{q}_i \rVert$ that measures how much the disks need to be moved to transform configuration $\config{p}$ into $\config{q}$ where $\vec{p}_i$ is the position of the $i$th disk in the fundamental cell \cite{ericok2021quotient}. If an equivalence relation derives from a set of isometries $\mathcal{S}$, then a corresponding metric $d_{\Lambda/\mathcal{S}}$ on the quotient space $\Lambda/\mathcal{S}$ can be written as \cite{burago2001course}
\begin{equation}
d_{\Lambda/\mathcal{S}}(\config{p},\config{q}) = \inf \limits_{\substack{S \in \mathcal{S}}}\{ d_\Lambda [\config{p},S(\config{q})] \}.
\label{eq:distance_quotient}
\end{equation} 
The computation of $d_{\Lambda/\mathcal{S}}$ is formulated as a global optimization problem and solved by the tabu search algorithm \cite{glover1989tabu,chelouah2000tabu}. After the pairwise distances $d_{\Lambda/\mathcal{S}}$ between the sampled configurations are computed, the ISOMAP algorithm is used to construct an approximate embedding into a Euclidean space. 

Both methods initially sample configurations of $n$ points in the RD, map them into a quotient configuration space, use the appropriate distance function to metrize this space, and triangulate the resulting point cloud as an $\alpha$-complex \cite{edelsbrunner1983shape}. While both methods give quotient spaces with the same topology, the second method better preserves the quotient spaces geometry.

Although the Eq.\ \ref{eq:distance_quotient} can be extended to the configuration spaces of hard spheres without difficulty, the computational cost of the algorithm is an issue. The computation of $d_{\Lambda/\mathcal{S}}$ involves a global optimization problem, and constructing the pairwise distance matrix of $m$ configurations for a naive implementation would require solving $\mathcal{O}(m^2)$ of these problems. Instead, the symmetry-invariant descriptors for configurations of hard spheres (derived in App.\ \ref{sec:app_descriptors}) are used in a preprocessing step to reduce the computational cost. Recall that the ISOMAP algorithm only requires the lengths of the edges of the $k$-nearest neighbor graphs as inputs. These graphs can be approximated using the distances between configurations as calculated by the descriptors, \ie the $k'$-nearest neighbors found using the descriptors should include all of the $k$-nearest neighbors that would be found using $d_{\Lambda/\mathcal{S}}$ for sufficiently large $k' > k$. The pairwise distances $d_{\Lambda/\mathcal{S}}$ then only need to be calculated for the $k'$-nearest neighbors, reducing the computational cost to $\mathcal{O}(m k')$. Finally, the desired $k$-nearest neighbor graphs with edge lengths determined by $d_{\Lambda/\mathcal{S}}$ are extracted and supplied to the ISOMAP algorithm.

One of the weaknesses of the above approach is that it does not reveal the minimum dimension of the Euclidean space required for the quotient map to be an embedding. For example, an embedding of the configuration space of two spheres requires more than six dimensions, but the ISOMAP algorithm does not protest when asked to construct an isometric embedding in four dimensions even though this results in self intersections.

\begin{figure}
	\centering
	\includegraphics[width=0.85\columnwidth]{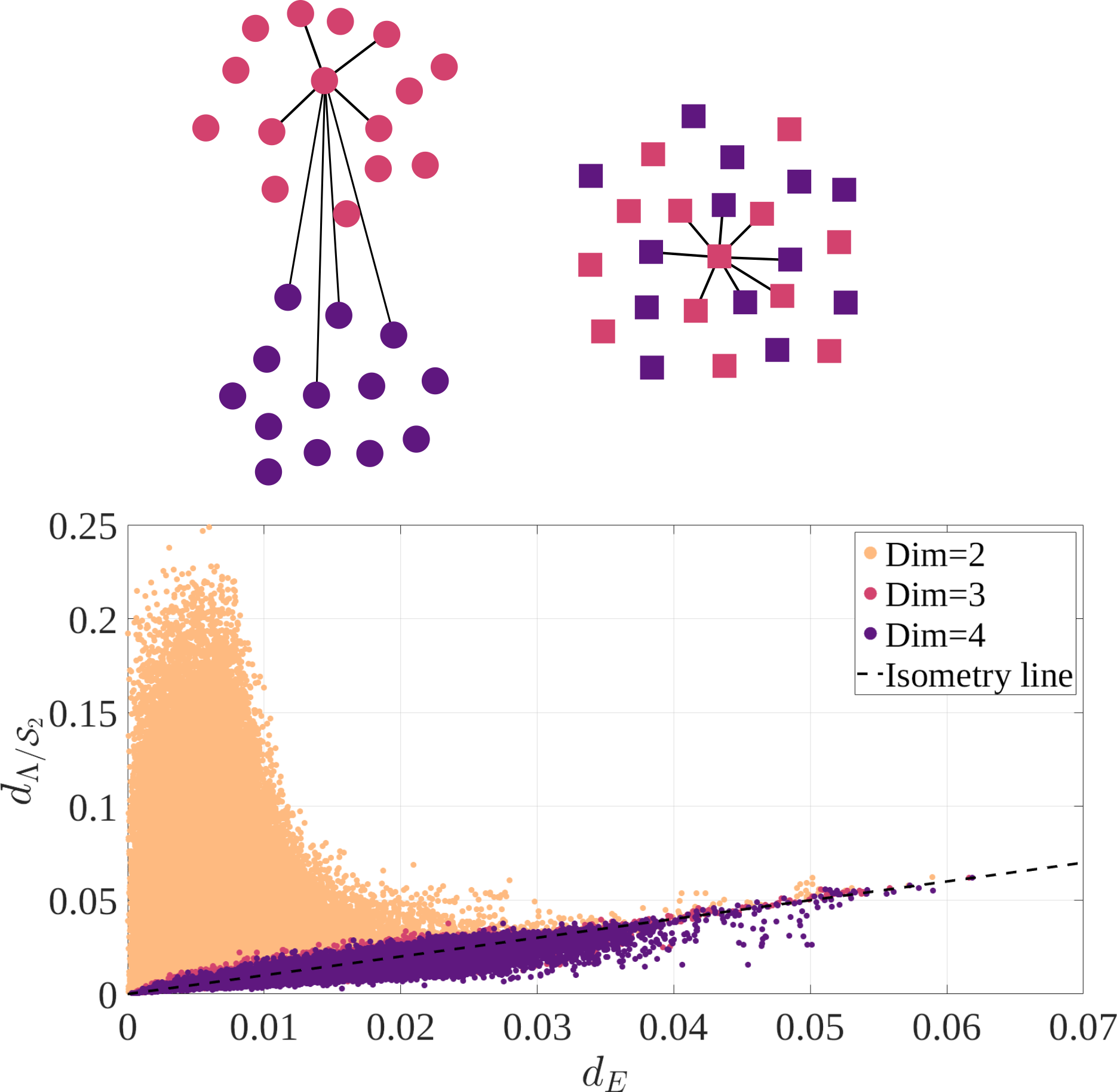}
	\caption{Top panel shows an improper embedding, \ie two separate regions in $\Lambda/\mathcal{S}$ (left) are merged in the new Euclidean space (right). Given a point in the new space, the pairwise distances to its neighbors are small whereas the corresponding pairwise distances in the original space are large. The bottom panel shows the result of this analysis for $\Lambda/\mathcal{S}_2$ for $n=2$ spheres.}
	\label{fig:inverse_dim_analysis}
\end{figure}

Whether the ISOMAP algorithm in fact generates an embedding is checked by the following procedure. The map of the quotient space into a Euclidean space should preserve the local geometry isometrically, \ie all the distances between a configuration and its $k$-nearest neighbors before and after the transformation should be preserved. This suggests that a simple scatter plot of the distances before and after the transformation could indicate something about the success of the operation. The top panel in Fig.\ \ref{fig:inverse_dim_analysis} represents a schematic for the inverse analysis. After the new coordinates (squares) are constructed, a $k$-nearest neighbor graph of each point in the new space is found, and the pairwise Euclidean distances $d_E$ in the new space are compared to the corresponding $d_{\Lambda/\mathcal{S}}$ in the old space. If the embedding is proper, then these distances should follow the isometry line with little distortion. If it is not, then some of the short edges in the new space will map back to long edges in the old space. The bottom panel shows the result of this analysis for $\Lambda/\mathcal{S}_2$ for $n=2$ spheres, and provides clear visual evidence that at least three coordinates are needed for a proper embedding of $\Lambda/\mathcal{S}_2$.

The ideas developed so far are applied to the configuration spaces of $n = 2$ spheres because it is the most complicated system that can still be visualized. The summary of the critical points sampled by the procedure in Sec.\ \ref{subsec:morse_theory} for the $n=2$ case are given in Table \ref{table:crits2}, and these configurations can be visualized using the online database. Following Ref.\ \cite{ericok2021quotient}, it is observed that the 0-radius configuration acts like an index-3 critical point, and is added to the list even though it is not listed as a critical point in the online database.

\begin{table}
	\centering
	\begin{tabular}{c c c c c} 
		Index & Radius & Packing fraction & \# in $\Lambda/\mathcal{S}_1$& \# in $\Lambda/\mathcal{S}_2	$\\ [0.5ex] 
		\hline\hline
		0	&0.3560	&0.5236	& 1	&1 \\
		0	&0.3062	&0.3401	& 2 &1 \\
		1	&0.2887	&0.2850	& 8 &1 \\
		2	&0.2500	&0.1851	& 6 &1 \\
		3	&0		&0		& 1 &1 \\[1ex]
	\end{tabular}
	\caption{The critical points found after millions of random initializations for $n = 2$. Observe that the number of critical points changes dramatically as the space is quotiented by an increasing number of symmetries.}
	\label{table:crits2}
\end{table}

\subsection{Adding translation invariance}
\label{subsec:config_t}

\begin{figure*}
	\centering
	\includegraphics[width=1.0\textwidth]{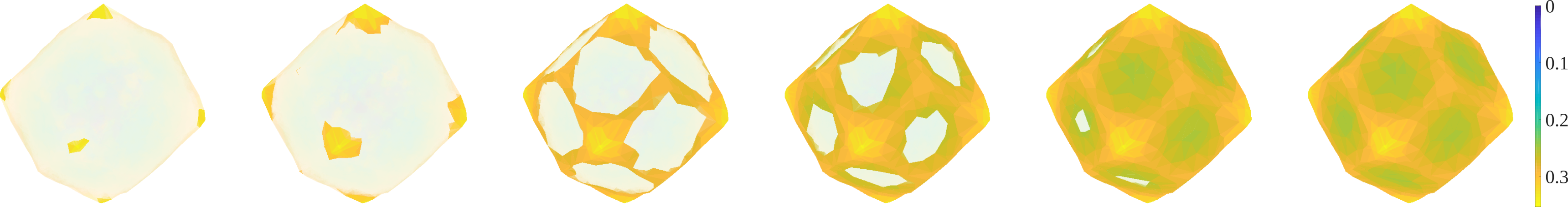}
	\caption{The evolution of the configuration space quotiented by translations $\Gamma(2, \rho) / \mathcal{S}_1$ (periodic boundary conditions are imposed by identifying opposite faces) with $\rho=\{0.31, 0.29, 0.27, 0.26, 0.25, 0.21\}$.}
	\label{fig:config_t}
\end{figure*}

For a system of $n=2$ spheres, fixing the first sphere at the origin effectively quotients the configuration space by translations. Since the full configuration space for $n=2$ is the product space $\mathbb{T}^3 \times \mathbb{T}^3$, the translation invariant configuration space $\Lambda/\mathcal{S}_1$ is in fact equivalent to a 3-dimensional torus $\mathbb{T}^3$ as expected. Figure \ref{fig:config_t} shows the evolution of this space as a function of sphere radius. $\Gamma(2,0.31)/\mathcal{S}_1$ is comprised of six 0-handles corresponding to the first index-0 critical point. It should be noted that since each of these handles is shared by six rhombic dodecahedra, there is only one index-0 critical point with $\rho=0.3560$ in this space. The 0-handles corresponding to the second index-0 critical point just appear in $\Gamma(2,0.29)/\mathcal{S}_1$. There are eight of these handles and they are each shared by four RD, resulting in two index-0 critical points with $\rho=0.3062$. Observe that for $\rho>0.2887$ the space is comprised of three disconnected handles, accounting for periodicity. These disconnected handles are connected by 24 1-handles corresponding to index-1 critical points, with each 1-handle shared by three RD for a total of eight index-1 critical points. The index-2 critical points appear in the middle of the rhombic faces when $\rho=0.25$ and are each shared by two RD for a total of six index-2 critical points. Finally, the space continues to fill for $\rho<0.25$, and the 0-radius configuration acts like an index-3 critical point that closes the space. The topology of $\Lambda/\mathcal{S}_1 = \Gamma(2,0)/\mathcal{S}_1$ is that of a 3-dimensional torus.

\subsection{Adding permutation and lattice invariance}
\label{subsec:config_ptil}

\begin{figure*}
	\centering
	\includegraphics[width=1.0\textwidth]{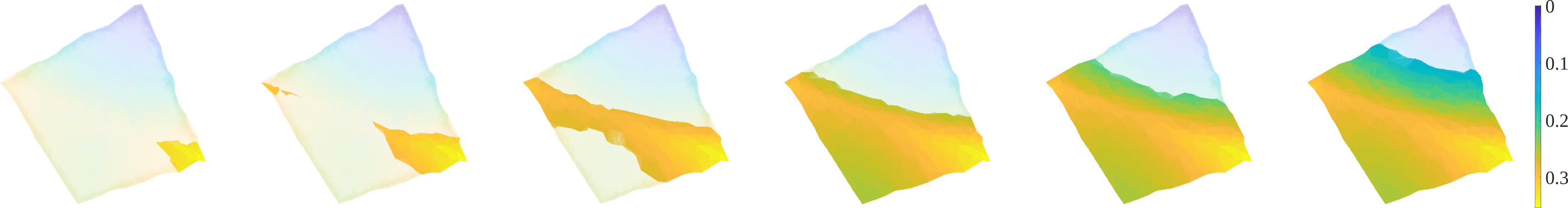}
	\caption{The evolution of the configuration space quotiented by translations, permutations, inversions and lattice symmetries $\Gamma(2, \rho) / \mathcal{S}_2$ with $\rho=\{0.31, 0.29, 0.27, 0.25, 0.21, 0.15\}$.}
	\label{fig:config_ptil}
\end{figure*}

Figure \ref{fig:inverse_dim_analysis} suggests that three coordinates are needed to represent the configuration space quotiented by translations, permutations, inversions, and lattice symmetries. The evolution of this space is shown in Fig.\ \ref{fig:config_ptil} as a function of sphere radius. Observe that the first index-0 critical point has already appeared when $\rho=0.31$, and $\Gamma(2,0.31)$ is comprised of a single connected component, the 0-handle. When $\rho = 0.29$, $\Gamma(2,0.29)$ has two 0-handles as disconnected components. The index-1 critical point connects these when $\rho=0.2887$, and all the spaces for $\rho<0.2887$ have a single connected component. Further decreasing $\rho$ causes the index-2 critical point to appear when $\rho=0.25$ and continues to fill the space.

\subsection{Paths in configuration space}
\label{subsec:paths}

Although the triangulation of a configuration space allows us to study its topological and geometric properties in detail, it is sometimes desirable to have a simpler representation. As previously mentioned, the merge tree summarizes how the connected components of a space merge and separate. While useful, this retains no information about the higher index critical points.

Consider a graph with vertices corresponding to the critical points. An edge is placed between two critical points if and only if there exists a path in the space from one to the other such that the value of $\tau$ along the path is monotone. This ensures that the path will always be within the superlevel set constrained by the radii of the given critical points, and the path goes directly from one critical point to the other.

Finding such a path is not an easy task. Fortunately, there are well-established algorithms to find minimum-energy paths (MEPs) in the chemical physics literature. The first step to find such a path is to construct the straight line path connecting the two critical points. This path is then relaxed using the zero-temperature string method \cite{weinan2002string,weinan2007simplified}, aligning the path with the gradient field of the potential energy (the nudged-elastic band method \cite{jonsson1998nudged,henkelman2000climbing} accomplishes the same task). The resulting graph is referred to as the MEP graph.

\begin{figure}
	\centering
	\includegraphics[width=0.6\columnwidth]{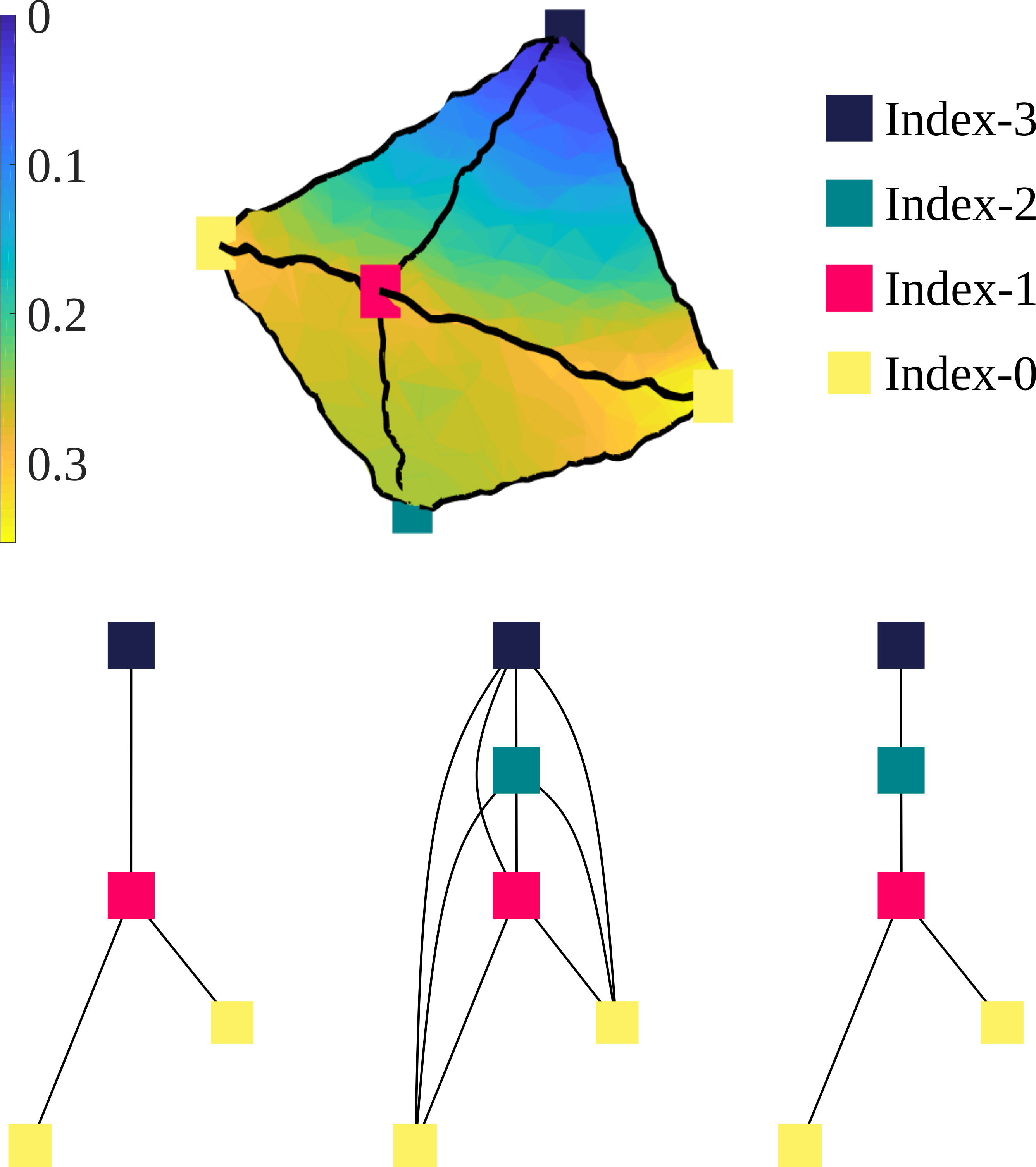}
	\caption{The top row shows the configuration space $\Lambda/\mathcal{S}_2$ quotiented by permutations, translations, inversions and symmetries of the tiling whereas the bottom row shows its merge tree (left), MEP graph (middle) and the sparsified MEP graph. The black curves on $\Lambda/\mathcal{S}_2$ are the actual MEPs connecting the critical points.}
	\label{fig:zts_merge}
\end{figure}

The top row of Fig.\ \ref{fig:zts_merge} shows the configuration space quotiented by permutations, translations, inversions and lattice symmetries $\Lambda/\mathcal{S}_2$. The MEPs between the critical points found with the procedure above are shown as black curves. Though not visible, there exists a path connecting the index-2 and index-3 critical points. The bottom of Fig.\ \ref{fig:zts_merge} shows the merge tree (left) and the MEP graph (middle); notice that the merge tree doesn't contain the index-2 critical point since the number of components doesn't change there, implying that the MEP graph in the middle contains strictly more information about the space. The two index-0 critical points are not connected by an edge because there is no way to go from one to the other along a path where $\tau$ is monotone. The MEP between them instead passes through the index-1 critical point, acting as a bridge to connect two disconnected portions of the space. This is exactly the location where significant changes to the topological and geometric properties are expected to occur. Finally, the graph on the bottom right in Fig.\ \ref{fig:zts_merge} is the sparsified MEP graph where only the edges between critical points with adjacent indices are included. This graph is visually simpler, and can be shown to include precisely the same information as the full MEP graph.

\section{Diffusion distance}
\label{sec:diffusion_distance}

Although the MEPs have a clear physical meaning, a thermodynamic system doesn't always follow these paths from one state to another. A concept like the diffusion distance instead effectively considers all the paths that the system could take, and provides a clearer connection to the mixing time of the system. 

\begin{figure*}
	\centering
	\includegraphics[width=0.75\textwidth]{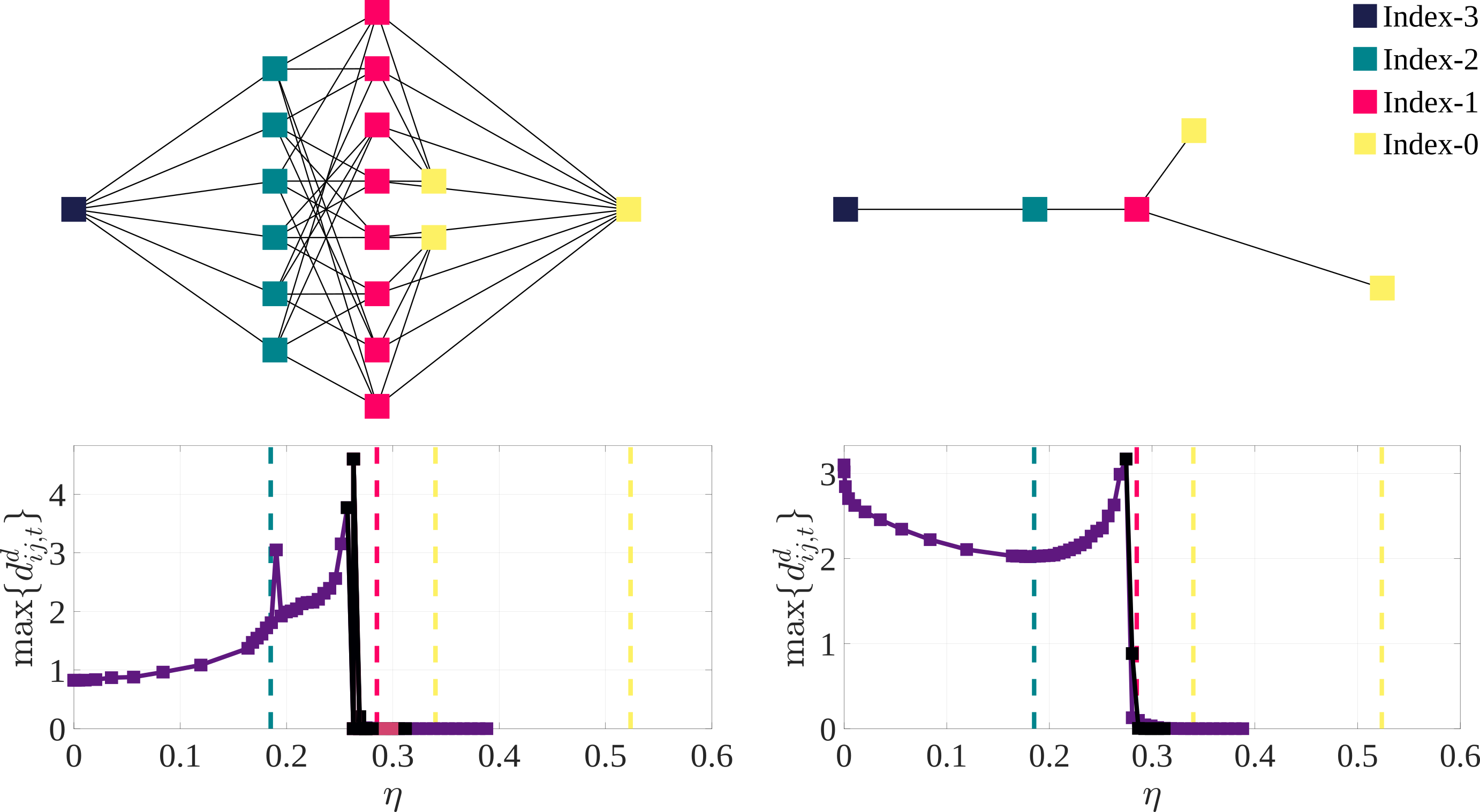}
	\caption{The sparsified MEP graphs of the configuration spaces quotiented by the translations (top left) and all the symmetries listed in Tab.\ \ref{table:symmetries} (top right). The corresponding diffusion distances at $t=8$ and $t=20$ are plotted below the sparsified MEP graphs. Observe that the discontinuous jumps coincide with the merging of the components; the small gap observed between the packing fraction of the critical points and the discontinuity in the left plot is due to finite resolution of the $\alpha$-complex.}
	\label{fig:cs_diff_dists}
\end{figure*}

The diameters of the configuration spaces $\Gamma(2,\eta)/\mathcal{S}_1$ (left) and $\Gamma(2,\eta)/\mathcal{S}_2$ (right) as measured by the diffusion distance are plotted on the bottom of Fig.\ \ref{fig:cs_diff_dists} as a function of packing fraction $\eta$. The top shows the sparsified MEP graphs of both spaces. The vertical lines indicate the corresponding packing densities of the critical configurations. In both cases, the disconnected regions of the space corresponding to index-0 critical configurations start growing slowly at the corresponding densities. When the index-1 critical configuration appears at $\eta=0.285$, the disconnected components get connected and discontinuous jumps in the diffusion distance are observed in both cases (the small gap in the left figure is due to the finite resolution of the $\alpha$-complex). From a thermodynamic standpoint, this is the location where the available portion of the configuration space and the domain of integration abruptly changes. Further decreasing the density decreases the diffusion distance since more paths are available to the system. Notice that the diffusion distance on the right starts to increase after the index-2 critical point appears. The zero density (or zero radius) configuration corresponds to a sharp corner, making it difficult for a random walker to diffuse into that region and increasing the diffusion distance.

Although the geometric analysis in this section is performed for $n = 2$ spheres, we conjecture that there will be a corresponding signature of a first-order phase transition around the melting packing fraction in the thermodynamic limit. The strongest support for this conjecture is the relationship of the critical point distribution (especially the index-$1$ critical points) to the coexistence interval in Fig.\ \ref{fig:distributions}. One of the main actions of index-$1$ critical points is to connect basins of attraction and thereby to introduce discontinuities in the space diameter. That the number and perhaps concentration of index-$1$ critical points within the coexistence interval both increase with $n$ suggests that there could be a catastrophic geometric event coinciding with the appearance of the liquid in the thermodynamic limit, and that this could be associated with a similarly catastrophic discontinuity in the space diameter. Many of the present results are intended to be the foundations of a future inquiry into whether this is indeed so.

\section{Conclusion}
\label{sec:conclusion}

The Topological Hypothesis claims that a change in the topology of the configuration space is necessary for a phase transition. Even though the original form of this hypothesis is no longer believed to be true, the distribution of critical points of the potential energy could still have a role in the onset of a phase transition. This work explores a new conjecture along these lines, \ie that a first-order phase transition occurs when there is a non-analyticity in the diameter of the configuration space as measured by the diffusion distance. This is believed to be closely related to the mixing time of the system.

This work develops a framework to collect evidence for the conjecture in the context of the configuration spaces of hard spheres. First, the critical points of the configuration spaces for $n=1 \dots 12$ spheres were sampled, tabulated in an online database, and made freely available as a Colab Notebook. Second, the definitions of the distance and the descriptors proposed in Ref.\ \cite{ericok2021quotient} were extended to hard spheres systems, and the configuration spaces with and without symmetry were explicitly triangulated as $\alpha$-complexes for a system of $n = 2$ spheres. Third, the minimum-energy paths between the critical configurations were found using the zero-temperature string method, and a new topological invariant called the minimum-energy graph was proposed. Finally, the diameters of the configuration spaces as measured by the diffusion distance were found for $n = 2$ spheres. The diffusion distance was found to be sensitive enough to capture the abrupt changes in the geometry of the spaces at the appropriate critical points. 

Future work will extend the diffusion distance analysis to a larger number of spheres to explore the conjecture that the discontinuities in space diameter already visible for the $n = 2$ system in Fig.\ \ref{fig:cs_diff_dists} are actually the signature of a catastrophic geometric event that drives the onset of a phase transition in the thermodynamic limit. The strongest evidence supporting this view is the distribution of lower index critical points in Fig.\ \ref{fig:distributions} which either remains centered in the middle of the coexistence interval or shifts closer to the packing fraction of the solid limit with increasing $n$. This implies a topological catastrophe at the melting density in the thermodynamic limit, with the corresponding geometric manifestation being a persistent discontinuity in the configuration space diameter. One major challenge would be the enumeration of all the critical points, since their number appears to increase exponentially with the number of spheres. A second could be the memory and computational cost required to construct the $\alpha$-complexes; perhaps the configuration spaces could be efficiently triangulated using sparsified Vietoris-Rips complexes.

\begin{acknowledgments}
O.B.E and J.K.M. were  supported  by  the  National  Science  Foundation under Grant No. 1839370.
\end{acknowledgments}

\appendix
\section{The 3-dimensional torus}
\label{sec:app_3-torus}
\begin{figure}
	\centering
	\includegraphics[width=0.6\columnwidth]{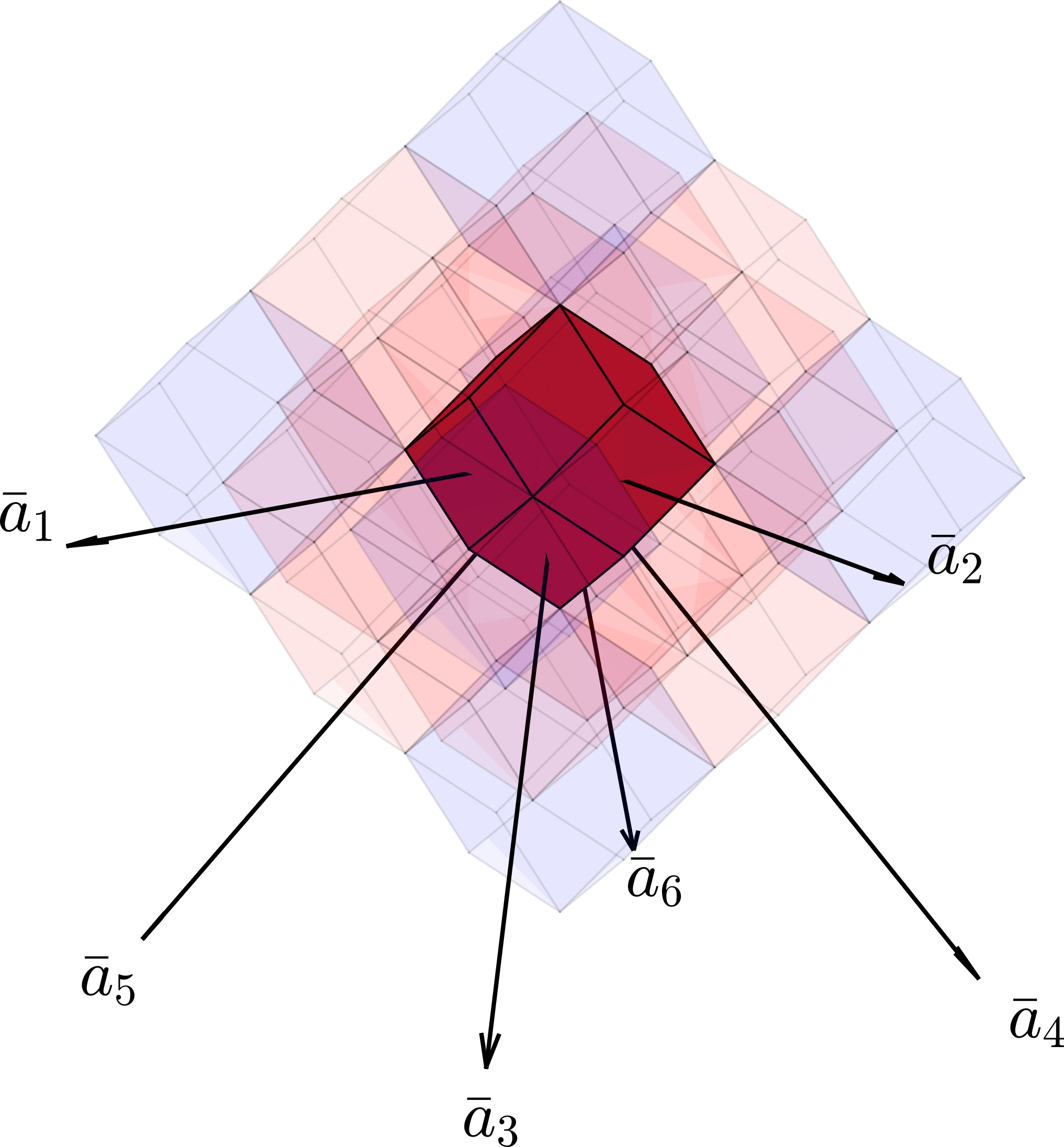}
	\caption{A 3-torus $\mathbb{T}^3$ is obtained by identifying opposite faces of a rhombic dodecahedron. The unit vectors $\vec{a}_1$ and $\vec{a}_2$ are respectively aligned with the $x$ and $y$ axes. The remaining four $\vec{a}_{i}$ pass through the centers of the lower faces.}
	\label{fig:rhombic_dodecahedron}
\end{figure}

This work studies the configuration spaces of hard spheres in a 3-dimensional torus $\mathbb{T}^3$. While $\mathbb{T}^3$ can be constructed by identifying opposite sides of various space-filling polyhedra, the rhombic dodecahedron (RD) with octahedral symmetry $O_h$ is used here; the identification of opposite faces means that an object that leaves by any of the twelve faces of the RD reenters by the opposite face. The justification for this choice is that one of the densest sphere packings in three dimensions can be constructed by repeated translations of the RD with its inscribed sphere. The dark red RD shown in Fig.\ \ref{fig:rhombic_dodecahedron} is the fundamental unit cell of this tessellation, and the faded cells are several of the periodic images. The twelve nearest neighbor cells share a face with the fundamental cell and are shown in faint red, whereas the six second-nearest neighbor cells share a vertex and are shown in faint blue. The center to center distance between nearest neighbor cells is one, which equivalently means that the volume of the fundamental cell is $\sqrt{2} / 2$.

\section{Simplicial complex}
\label{sec:app_simplicial_complex}

A computationally convenient way to study the topological and geometric properties of a space is to construct an explicit triangulation by means of a simplicial complex. A $k$-simplex is the generalization of a triangle to arbitrary dimensions, \ie a $0$-simplex is a point, a $1$-simplex is a line, a $2$-simplex is a triangle, etc. More formally, a $k$-simplex $\Delta_k$ is the convex hull of $k + 1$ affinely independent points in $\mathbb{R}^d$ for $0 \leq k \leq d$. A simplicial complex is roughly a collection of such simplices without improper intersections \cite{munkres2018elements}. There are various protocols to construct simplicial complexes like the Vietoris-Rips complex \cite{vietoris1927hoheren} or the Cech complex \cite{hatcher2002algebraic} from a set of points. This work uses the $\alpha$-complex \cite{edelsbrunner1983shape} which is a subcomplex of the Delaunay triangulation \cite{delaunay1934sphere}. Formally, let $DT(X)$ be the Delaunay triangulation of a point cloud $X$ in $\mathbb{R}^d$. For a given $\alpha$, the $\alpha$-complex $C_{\alpha}(X)$ of $X$ is the subcomplex containing all simplices of $DT(X)$ that have circumradii less than or equal to $\alpha$.

\begin{figure}
	\centering
	\includegraphics[width=1.0\columnwidth]{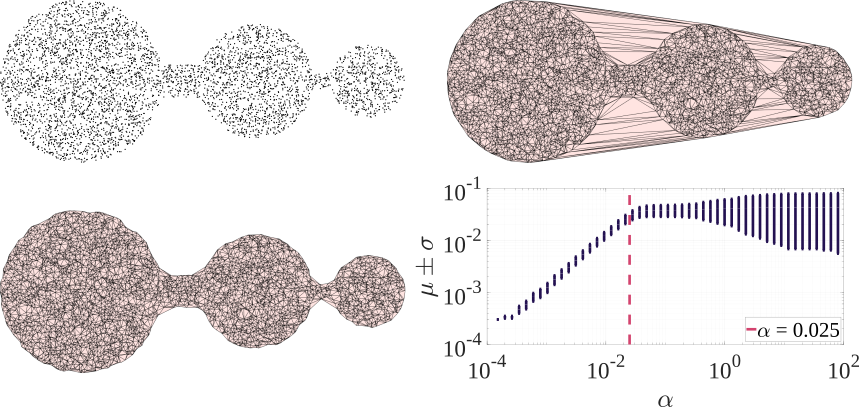}
	\caption{A schematic showing the construction of the $\alpha$-complex. The point cloud in the top left is comprised of points that are uniformly sampled inside the three circles of radius $1$, $0.7$ and $0.45$, and has the Delaunay triangulation on the top right. The proposed length scale analysis is on the bottom right, and the $\alpha$-complex with $\alpha = 0.025$ is on the bottom left.}
	\label{fig:alpha_complex_example}
\end{figure}

One important question is to how to choose the value of $\alpha$ such that $C_{\alpha}(X)$ adequately reflects the geometry of the underlying manifold. While the standard approach in the literature involves persistent homology \cite{edelsbrunner2008persistent}, a simple length scale analysis is likely sufficient for our purposes. Let $\mu$ be the average length of an edge in the $\alpha$-complex and $\sigma$ be the standard deviation of the edge lengths. When $\alpha$ is very small, the complex contains many disconnected components and poorly represents the geometry of the underlying manifold. For these values, $\mu$ increases with $\alpha$ as successively longer edges appear. When $\alpha$ is very large, the complex contains simplices that extend outside of the underlying manifold and does not adequately constrain paths through the space. For these values, inclusion of the longest edges increases $\sigma$ substantially. There is generally an interval of intermediate $\alpha$ where $\mu$ is relatively stable and $\sigma$ has not yet increased for which the complex is a good approximation to the underlying manifold. Figure \ref{fig:alpha_complex_example} shows the procedure to construct the $\alpha$-complex for an example point cloud along with the proposed length scale analysis.

\begin{figure*}
	\centering
	\includegraphics[width=1.0\textwidth]{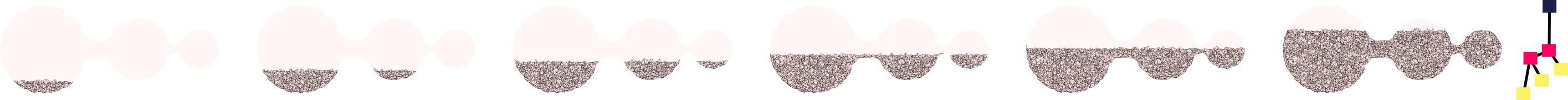}
	\caption{The evolution of the example $\alpha$-complex by the height filtration. Observe that the topology only changes when the filtration reaches to the heights of the channels. The merge tree of the $\alpha$-complex is shown on the right where the yellow squares are the local minima, the pink squares are saddles, and the dark blue is the global maximum.}
	\label{fig:example_filtrations}
\end{figure*}

Figure \ref{fig:example_filtrations} shows filtrations of the space in Fig.\ \ref{fig:alpha_complex_example} by the height function with $-1 < y < 1$. For $y < -1$ the $y$-sublevel set is empty. As $y$ is increased above $-1$ the bottom part of the largest disk appears first, with the bottom parts of the second and third disks following as $y$ increases further. When the $y$-sublevel set reaches the bottom of the first channel, a narrow channel appears that merges the first and second components. Further increasing $y$ causes this channel to slowly grow, and when $y$ reaches the bottom of the second channel all of the components merge into a single component. Higher values of $y$ fill out the space further but no longer change the topology.  The merge tree of the example $\alpha$-complex is shown on the right in Fig.\ \ref{fig:example_filtrations}.

\section{Commute time distance and diffusion distance}
\label{sec:app_graph_distances}

Given a weighted graph $G$, the commute time distance and the diffusion distance require the choice of a kernel $k_{ij}$ that provides a notion of similarity between vertices $i$ and $j$, with large values indicating that the points are close in the underlying manifold. The kernel should be (i) strictly positive such that $k_{ij} \geq 0$, and (ii) symmetric such that $k_{ij} = k_{ji}$. This work uses the standard Gaussian kernel $k_{ij} \propto \exp[-s_{ij}^2 / (2\sigma^2)]$ where $s_{ij}$ is the geodesic distance between vertices $i$ and $j$ and $\sigma$ is a length scale representing the locality of the neighborhood; a small $\sigma$ means that only the interactions between the closest neighbors significantly contribute. Unless otherwise specified, $\sigma$ is equal to the radius of a sphere with volume equal to $25\%$ that of the $\alpha$-complex.

\begin{figure}
	\centering
	\includegraphics[width=0.95\columnwidth]{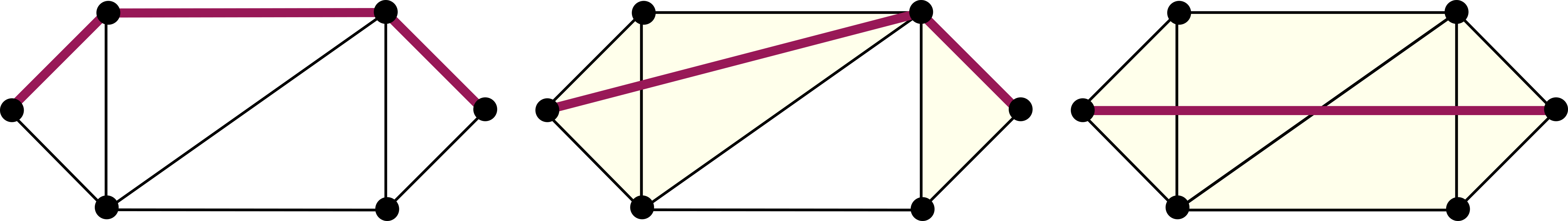}
	\caption{Example simplicial complexes (black points, black lines, and yellow triangles) and shortest paths (bold purple lines). The left complex consists of six $0$-simplices and nine $1$-simplices. The middle complex additionally has three $2$-simplices (yellow triangles). The right complex has one more $2$-simplex. The shortest path length between two vertices can decrease when higher-dimensional simplices are included.}
	\label{fig:example_simp_dist}
\end{figure}

While both distances are typically defined on graphs, they can easily be extended to higher-dimensional simplicial complexes. The essential difference is that a random walker on a graph is only allowed to travel along the edges of the graph, whereas a walker in a general simplicial complex is allowed to travel through the interiors of the simplices. The potential significance of this is clear if one considers a unit $n$-dimensional cube; a path between opposite corners along the cube edges is length $n$, whereas a path along the body diagonal is length $\sqrt{n}$. This work calculates the geodesic distances between vertices along paths allowed to pass through the interiors of simplices using the algorithm proposed by Bhattacharya \cite{bhattacharya2019towards}. The algorithm is effectively a modification of Dijsktra's algorithm, adding an additional step that allows paths to pass through the interior of a simplex with only one vertex whose distance from a central vertex is unknown. Figure \ref{fig:example_simp_dist} shows several examples of the shortest paths between points as calculated by this method. On the left, the interiors of the triangles are not included and the simplicial complex is only comprised of vertices and edges. The distance between any two vertices can be found by Dijkstra's algorithm \cite{dijkstra1959note}. Triangle interiors shown in yellow are included in the simplicial complexes in the middle and the right, allowing shorter paths through the space.

The calculation of the commute time distance begins with the $n \times n$ adjacency matrix $\mat{A}$ with elements $a_{ij}$ equal to $k_{ij}$ if $i$ and $j$ are in the same connected component and $0$ otherwise. The graph Laplacian $\mat{L}$ is defined as $\mat{L} = \mat{D} - \mat{A}$ where $\mat{D}$ is a diagonal matrix with elements $d_{ii} = \sum_{j} a_{ij}$. Let the elements of the Moore-Penrose inverse $\mat{L}^{+}$ be indicated by $l^{+}_{ij}$ \cite{rao1972generalized}. The commute time distance between a pair of vertices $d^{c}_{ij}$ is defined as \cite{fouss2006novel}
\begin{equation*}
d^{c}_{ij} = V_{G} \left(l^{+}_{ii}+l^{+}_{jj}-2l^{+}_{ij}\right)
\label{eq:commute_time_distance}
\end{equation*}
where $V_{G} = \sum_{k} d_{kk}$ is the volume of the graph.

The diffusion distance instead starts from the kernel matrix $\mat{K}$, containing the kernel values $k_{ij}$ for every pair of vertices $i$ and $j$. The transition matrix $\mat{P}$ is then defined as $\mat{P} = \mat{D}^{-1} \mat{K}$ where $\mat{D}$ is a diagonal matrix with elements $d_{ii} = \sum_j k_{ij}$. Intuitively, $\mat{P}$ contains the transition probabilities for a Markov chain process, and propagating this process forward by $t$ steps is equivalent to raising $\mat{P}$ to the $t^\mathrm{th}$ power. Rather than calculating the matrix power explicitly though, let $\lambda_l$ and $\vec{\phi}_l$ be the eigenvalues and eigenvectors of $\mat{P}$ for $l \ge 0$. Since $\mat{P}$ is normalized, the largest eigenvalue $\lambda_0$ is always $1$ with a constant eigenvector $\vec{\phi}_0$. Discarding the first eigenvalue and eigenvector, the diffusion coordinates are defined using $(n - 1)$ of the eigenvalues and eigenvectors as
\begin{equation*}
\vec{\Phi}_t(i) = \left[\lambda_1^t \vec{\phi}_1(i), \dots, \lambda_{n - 1}^t \vec{\phi}_{n - 1}(i)\right].
\label{eq:diffusion_maps}
\end{equation*}
Finally, the diffusion distance $d^{d}_{ij,t}$ at time step $t$ is defined as
\begin{equation}
d^{d}_{ij,t} = \norm{ \vec{\Phi}_t(i) - \vec{\Phi}_t(j) }
\label{eq:diffusion_distance}
\end{equation}
where $\norm{\cdot}$ is the Euclidean norm. The time step $t$ is usually chosen here to make the product $\sigma t$ approximately equal to the length of the shortest path between the two most distant points in the complete space.

\begin{figure}[h]
	\centering
	\includegraphics[width=0.7\columnwidth]{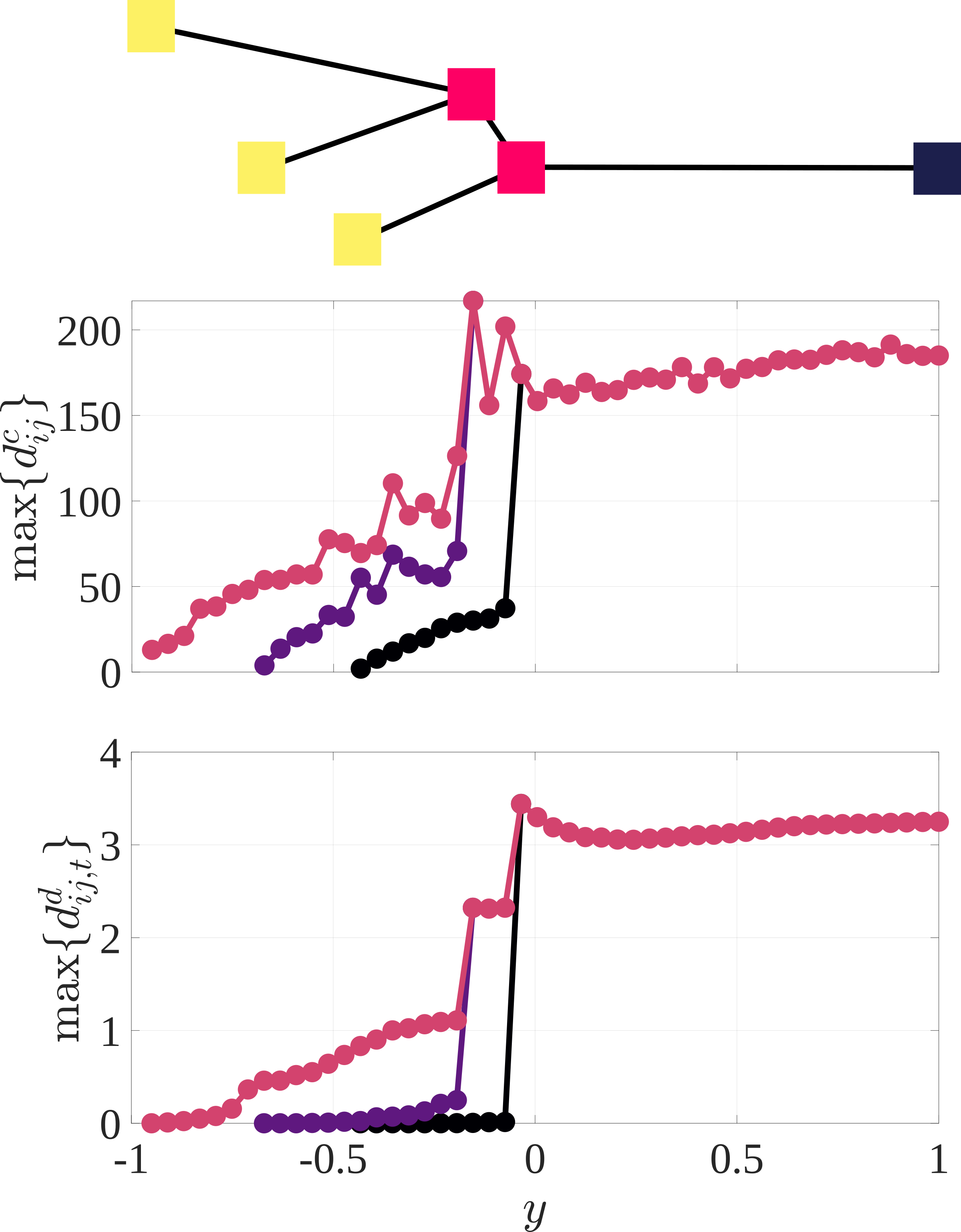}
	\caption{The diameter of the sublevel sets as measured by the commute time distance (top) and the diffusion distance (bottom) at $t = 50$ for the filtration in Fig.\ \ref{fig:example_filtrations}. Point color is associated with a particular component of the space. The discontinuities in the diameter coincide with the merging of components.}
	\label{fig:example_graph_dists}
\end{figure}

Figure \ref{fig:example_graph_dists} shows the diameter of the spaces in Fig.\ \ref{fig:example_filtrations} as a function of $y$ for the commute time distance (middle) and the diffusion distance (bottom). The similarity kernel $\mat{K}$ is the standard Gaussian kernel with a standard deviation of $0.1$. Observe that the three disconnected components appear as $y$ is increased. The discontinuous jumps in the diffusion diameter (and to a lesser extent in the commute time diameter) correspond to the sublevel set values for which previously disconnected components of the space merge, making the accessible region of the space larger. These are precisely the locations where the topology of the space changes, \ie the saddle points marked as pink squares in the merge tree on the top of Fig.\ \ref{fig:example_graph_dists}. This provides preliminary evidence that the diffusion distance could be a reasonable choice for the distance in Conjecture \ref{conj:phase_transition}; changes in the value of the diffusion diameter appear to indicate corresponding changes in the topology and the extent of the accessible region, and therefore in the mixing time of a thermodynamic system. While the index-$1$ critical points (saddles) can be thought of as channels connecting index-$0$ critical points (minima), the effect of the higher index critical points is more complicated. Recalling the discussion of the distance between cube corners above, an index-$k$ critical point could correspond to being able to pass through the diagonal of a $k$-dimensional cube, providing shortcuts between previously distant regions.

\section{Descriptors}
\label{sec:app_descriptors}

This section extends the definitions of the descriptors proposed in Ref.\ \cite{ericok2021quotient} to the hard sphere model. Consider $n$ hard spheres in three dimensional torus $\mathbb{T}^3$. The derivation starts by defining a new density function $f(\vec{a})$ as the sum of Dirac-delta distributions located at the sphere centers $\vec{a}_j$ with respect to the $a_1 a_2 a_3$-coordinate system shown in Fig.\ \ref{fig:rhombic_dodecahedron}, or
\begin{equation*}
f(\vec{a}) = \sum_{j=1}^{n} \delta(\vec{a}_j).
\label{eq:app_density_function}
\end{equation*}

\noindent Observe that the density function is invariant under the permutations of the sphere labels due to the commutative property of summation. Expanding $f(\vec{a})$ as a Fourier series yields
\begin{equation*}
f(\vec{a}) = \sum_{\vec{k}} c_{\vec{k}} \mathrm{e}^{2 \pi \mathrm{i} \vec{k} \cdot \vec{a}}
\label{eq:app_fourier_series}
\end{equation*}
where $c_{\vec{k}}$ are the complex coefficients of the expansion and $\vec{k} = [ u,v,w ]$ for integers $u$, $v$ and $w$. By orthogonality of the complex exponentials, the infinite set of coefficients can be calculated as
\begin{equation*}
c_{\vec{k}} = \sum_{j=1}^{n} \mathrm{e}^{-2 \pi \mathrm{i} \vec{k} \cdot \vec{a}_j}.
\label{eq:app_coeffs}
\end{equation*}
The complex coefficients respect the periodicity of the lattice. Moreover, translating the configurations as rigid bodies only changes the phase of the coefficients, keeping their moduli constant. Hence, a new set of real-valued descriptors that are additionally invariant under translations is defined as
\begin{equation*}
z_{\vec{k}} = \sqrt{c^\ast_{\vec{k}} c_{\vec{k}}}
\label{eq:app_desc_pti}
\end{equation*}
where ${}^{\ast}$ is the complex conjugate. It is easy to show that $z_{\vec{k}}$ is also invariant under the inversion symmetry. Finally, descriptors invariant under the lattice symmetries (octahedral symmetry $O_h$) can be defined as
\begin{equation*}
\hat{z}_{\vec{k}} = \frac{1}{48}\sum_{t=1}^{48} z_{\vec{k}}^{t}
\label{eq:app_desc_ptil}
\end{equation*}
where $z_{\vec{k}}^{t}$ are the descriptors $z_{\vec{k}}$ calculated for the configuration generated by applying the $t$th symmetry element in $O_h$ to the configuration. 

Not all $\hat{z}_{\vec{k}}$ are independent. Let $\vec{x}$ and $\vec{a}$ denote the coordinates of a given point in the $xyz$- and $a_1a_2a_3$-coordinate systems, respectively. The relationship of the two coordinate systems is given by the transformation matrix $\mat{T}$, \ie $\vec{a} = \mat{T} \vec{x}$ and $\vec{x} = \mat{T}^{-1} \vec{a}$. Let $\mat{O} \in O_h$ be a symmetry element written as a matrix in the $xyz$-coordinate system. A symmetric copy $\vec{x}'$ of a vector $\vec{x}$ can be written as $\vec{x}' = \mat{O} \vec{x}$. Then, 
\begin{equation*} 
\vec{a}' = \mat{T} \vec{x}' = \mat{T} \mat{O} \vec{x} = \mat{T} \mat{O} \mat{T}^{-1} \vec{a} = \mat{U} \vec{a}  
\end{equation*}
where $\mat{U} = \mat{T} \mat{O} \mat{T}^{-1}$. Observe that $\mat{U}$ corresponds to the symmetry matrix $\mat{O}$ written in the $a_1a_2a_3$-coordinate system. 

Suppose we compute $z_{\vec{k}}^{t}$ for the $t$th symmetry element in $O_h$. Then,
\begin{equation*} 
z_{\vec{k}}^{t} = \bigg\Vert \sum_{j = 1}^{n} \exp{(-2\pi \mathrm{i} 
	\vec{k} \cdot \mat{U}_t \vec{a}_j)} \bigg\Vert = \bigg\Vert \sum_{j = 1}^{n} \exp{(-2\pi \mathrm{i} 
	\vec{k}' \cdot \vec{a}_j)} \bigg\Vert
\end{equation*}
where $\vec{k}' = \vec{k} \cdot \mat{U}_t$. Computing $\vec{k}'$ for all elements of $O_h$ yields the equivalent indices presented in Tab. \ref{table:indices}.

\begin{table*}
	\centering
	\begin{tabularx}{0.8\textwidth}{c c c || c c c || c c c} 
		$k'_1$ & $k'_2$ & $k'_3$ & $k'_1$ & $k'_2$ & $k'_3$& $k'_1$ & $k'_2$ & $k'_3$\\ [0.5ex] 
		\hline\hline
		$u$	& $v$ & $w$	& $w-v$ & $u-w$ & $u$ & $w-v-u$	& $w$ & $w-u$	\\
		$u+v-w$ & $-w$ & $v-w$ & $u-w$	& $w-v$ & $-v$	& $v$ & $-u$ & $v-w$ \\
		$-w$	& $u+v-w$ & $u-w$	& $w$ & $w-v-u$ & $w-v$ & $w$	& $u+v-w$ & $v$	\\
		$u$ & $-v$ & $u-w$ & $w-v-u$	& $-w$ & $-v$	& $-v$ & $u$ & $u-w$ \\
		$w-u$	& $w-v$ & $w-v-u$	& $u$ & $v$ & $u+v-w$ & $w-v$	& $u-w$ & $-v$ \\
		$w-v-u$ & $w$ & $w-v$ & $v$	& $u$ & $w$	& $w-u$ & $v-w$ & $-u$ \\
		$u-w$	& $v-w$ & $-w$	& $w$ & $u+v-w$ & $u$ & $u$	& $-v$ & $w-v$\\
		$w-v$ & $w-u$ & $w$ & $-u$ & $v$ & $w-u$	& $v-w$ & $u-w$ & $u+v-w$ \\
		$-u$	& $-v$ & $-w$	& $v-w$ & $w-u$ & $-u$ & $u+v-w$	& $-w$ & $u-w$	\\
		$-v$ & $-u$ & $w-v-u$ & $w-u$	& $v-w$ & $v$	& $u-w$ & $w-v$ & $u$ \\
		$-u$	& $v$ & $v-w$	& $-w$ & $u+v-w$ & $v-w$ & $-w$	& $w-v-u$ & $-v$ \\
		$w-v-u$ & $-w$ & $-u$ & $u+v-w$	& $w$ & $v$	& $v$ & $-u$ & $w-u$ \\
		$u-w$	& $v-w$ & $u+v-w$	& $-u$ & $-v$ & $w-v-u$ & $v-w$	& $w-u$ & $v$	\\
		$v$ & $u$ & $u+v-w$ & $-v$	& $-u$ & $-w$	& $-v$ & $u$ & $w-v$ \\
		$w-u$	& $w-v$ & $w$	& $-w$ & $w-v-u$ & $-u$ & $w$	& $w-v-u$ & $w-u$\\ $v-w$ & $u-w$ & $-w$ & $u+v-w$	& $w$ & $u$	& $w-v,w-u,$ & $w-u$ & $w-v-u$ \\[1ex]
	\end{tabularx}
	\caption{The equivalent indices where $k'_i$ denotes the components of $\vec{k}'$.}
	\label{table:indices}
\end{table*}

\end{document}